\documentclass[acmsmall]{acmart}

\usepackage{hyperref}
\usepackage{color}
\usepackage{multirow} 
\usepackage[ruled,linesnumbered]{algorithm2e}

\AtBeginDocument{%
  \providecommand\BibTeX{{%
    \normalfont B\kern-0.5em{\scshape i\kern-0.25em b}\kern-0.8em\TeX}}}

\begin{document}

\title{LightFR: Lightweight Federated Recommendation with Privacy-preserving Matrix Factorization}

\author{Honglei Zhang}
\email{honglei.zhang@bjtu.edu.cn}
\affiliation{
  \institution{Beijing Jiaotong University}
  \city{Beijing}
  \country{China}
}
\author{Fangyuan Luo}
\email{fangyuanluo@bjtu.edu.cn}
\affiliation{
  \institution{Beijing Jiaotong University}
  \city{Beijing}
  \country{China}
}
\author{Jun Wu}
\email{wuj@bjtu.edu.cn}
\affiliation{
  \institution{Beijing Jiaotong University}
  \city{Beijing}
  \country{China}
}
\author{Xiangnan He}
\email{hexn@ustc.edu.cn}
\affiliation{
  \institution{University of Science and Technology of China}
  \city{Hefei}
  \country{China}
}
\author{Yidong Li}
\authornote{Corresponding author.}
\affiliation{
  \institution{Beijing Jiaotong University}
  \city{Beijing}
  \country{China}
}
\email{ydli@bjtu.edu.cn}

\renewcommand{\shortauthors}{Honglei Zhang, et al.}

\begin{abstract}
Federated recommender system (FRS), which enables many local devices to train a shared model jointly without transmitting local raw data, has become a prevalent recommendation paradigm with privacy-preserving advantages. However, previous work on FRS performs similarity search via inner product in continuous embedding space, which causes an efficiency bottleneck when the scale of items is extremely large. We argue that such a scheme in federated settings ignores the limited capacities in resource-constrained user devices (\textit{i.e.}, storage space, computational overhead, and communication bandwidth), and makes it harder to be deployed in large-scale recommender systems. Besides, it has been shown that transmitting local gradients in real-valued form between server and clients may leak users' private information. To this end, we propose a lightweight federated recommendation framework with privacy-preserving matrix factorization, \textit{LightFR}, that is able to generate high-quality binary codes by exploiting learning to hash technique under federated settings, and thus enjoys both fast online inference and economic memory consumption. Moreover, we devise an efficient federated discrete optimization algorithm to collaboratively train model parameters between the server and clients, which can effectively prevent real-valued gradient attacks from malicious parties. Through extensive experiments on four real-world datasets, we show that our LightFR model outperforms several state-of-the-art FRS methods in terms of recommendation accuracy, inference efficiency and data privacy.
\end{abstract}

\ccsdesc[500]{Information systems~Collaborative filtering}
\ccsdesc[500]{Security and privacy~Privacy protections}

\keywords{Federated Recommender System, Matrix Factorization, Privacy Preservation, Learning to Hash}

\maketitle

\section{Introduction}

Recommender system (RS) is an effective functionality for alleviating information overload~\cite{cf_survey_2014}, with the rapid growth of online user interaction data. The significance of RS cannot be overstated, regarding their widespread utilization in industry, such as web search and e-commerce platforms~\cite{youtubenet_2016,web_search_2021}, and their potential to surmount obstacles with user modeling in academia~\cite{cf_survey_2014,tseq_2020}. However, such a scheme that all behavior data is collected in a centralized manner, will inevitably result in the leakage of private user information~\cite{privacy_leakage_2008,deep_leakage_2019}. Thus, privacy concerns in RS arise. Considering the sensitivity of user personal data, regulations such as General Data Protection Regulation (GDPR)\footnote{https://gdpr-info.eu/}, have been put into effect to restrict the centralized collection of users' private data. Such actions lead to the occurrence of data isolation trend, which aggravates the data sparsity issue in RS scenarios.

Focusing on this dilemma, federated recommender system (FRS) has received widespread attention~\cite{frs_survey_2020}, due to the advantages of privacy protection and considerable performance. In FRS, a global model in the server can be aggregated and updated from user-specific local models with the collaboration of the server and clients, ensuring that users' private interaction data never leaves their devices. Among them, the more prominent work is Federated Collaborative Filtering (FCF)~\cite{fcf_2019}, where each user latent vector is updated locally and the item latent matrix is transmitted and updated collaboratively between the server and clients. Subsequently, FedFast~\cite{fedfast_2020} improves the client sampling strategy and the active aggregation mechanism on the shoulders of FCF, and speeds up the convergence efficiency while guaranteeing the model efficacy. Although the above methods realize high-quality federated recommendations, it requires transmitting a full amount of item latent matrix between the server and clients, which brings a huge carbon footprint and is unaffordable for cost-conscious clients.

Following that, some work explores reducing the payload of the entire item latent matrix by meta learning techniques~\cite{metamf_2020,privrec_2021}. For example, MetaMF~\cite{metamf_2020} adopts the meta recommender to deploy smaller models on the client to reduce memory consumption and PrivRec~\cite{privrec_2021} introduces a first-order meta-learning model to enable fast adaptation on local devices. Besides, there are also some attempts to utilize knowledge distillation mechanisms concentrating on transferring knowledge from a heavy model to a light one to achieve the purpose of designing lightweight recommender models on clients~\cite{llrec_2020,fedkd_2022}. Whereas these methods alleviate the storage and communication overheads to some extent, they can not shorten the inference time on clients since they should perform forward propagation in dense embedding space, which is unacceptable on computing-sensitive clients when the number of candidates is quite large. Besides, such solutions need to transmit the original real-valued gradient data, and it has been proved that the transmission of local gradients between the server and clients in continuous embedding space may leak users’ private information~\cite{fedmf_2020}. Consequently, it is a crucial challenge for existing FRS techniques to enhance the capability of preserving users' privacy.

\begin{figure}[t]
\begin{center}
\includegraphics[scale=0.7]{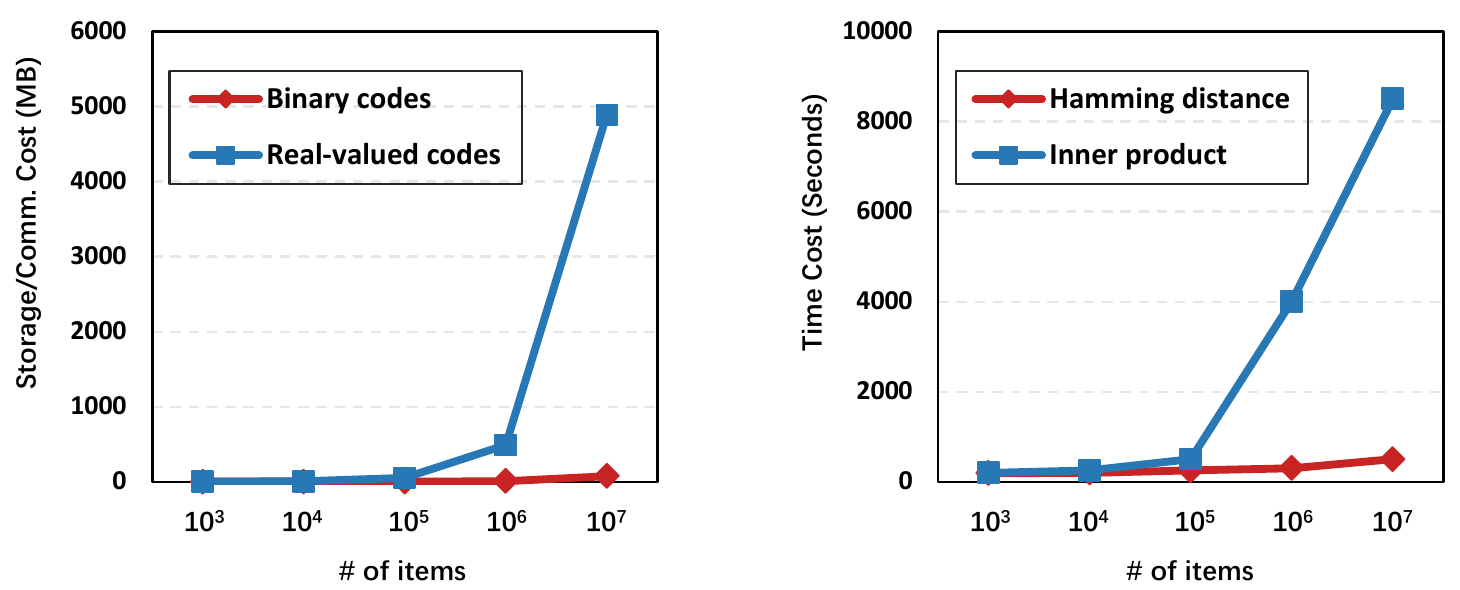}
\end{center}
\caption{Comparisons on storage, communication cost (on left panel) and inference time (on right panel) for Hamming distances (binary codes) and inner products (real-valued codes) on various-scaled items. The experiments are conducted by randomly generating binary codes and real-valued vectors with a length of 64 on $10^5$ users and $10^3-10^7$ items, and we report the average results over 5 repetitions\protect\footnotemark.
Since the number of items has a wide range, we display the x-axis evenly in the form of exponential intervals.
}
\label{fig:stcomp}
\end{figure}
\footnotetext{We perform the similarity search on a machine with 2.0GHz Intel Xeon E5-2640 processor.}

Along with this research line, several endeavors are devoted to facilitating the privacy of FRS~\cite{fedrec_2020,fedmf_2020}. For instance, FedRec~\cite{fedrec_2020} employs a hybrid filling strategy to randomly sample some virtual items to protect the actual gradients, and FMF-LDP~\cite{fmf-ldp_2021} adopts the differential privacy mechanism and a proxy network to reduce  the fingerprint surface for implicit data. Although these mechanisms can protect users' privacy to a certain level, they require huge memory, calculation and communication overheads, which are often not applicable to resource-constrained clients under federated settings. We argue that previous work on FRS generally provides top-K recommendations among all existing items via inner product in dense vector space, which is the leading factor for the challenge of high cost of resources in large-scale recommendation scenarios. Intuitive results can be observed from Fig.~\ref{fig:stcomp}, which shows that as the number of items increases, the storage cost, communication overhead, and inference time on the user terminal devices will rise dramatically. Note that different from centralized RS in a data server, FRS that requires to be deployed on local clients with low-resource settings, has more stringent restrictions for model scale. 
Hence, a lightweight recommendation model is even more urgent in FRS. In summary, none of these methods take into account both the issues of \textit{efficiency} and \textit{privacy}, which are the two primary challenges for real-world FRS.

Considering the shortcomings of existing work, we believe it is essential to develop a lightweight and privacy-preserving FRS, which not only benefits from the low cost of resources, but also increases the capability of privacy protection. To achieve this, we resort to the notion of learning to hash to obtain the binary representations of users and items so that the efficiency and privacy issues can be effectively addressed. However, solving the discrete optimization in federated settings is not trivial, since it is infeasible to utilize the straightforward heuristics because it is generally an NP-hard problem which involves exponential combinatorial searches for the binary codes. Hence, it is imperative to design an efficient federated discrete algorithm between the server and clients, which can embed the preferences of users into the discrete Hamming space, and meanwhile reduce resource utilization on both the server and clients in a privacy-persevering manner.

To remedy these issues in a unified way, we present \emph{LightFR}, a principled lightweight and privacy-preserving framework for FRS built on learning to hash and federated learning techniques. Specifically, we introduce a federated discrete optimization algorithm that can solve the above-mentioned issue in a computationally tractable fashion in federated settings, which can produce suitable binary user representation on local clients and binary item representations on the server side. By introducing learning to hash into LightFR, it can kill three birds with one stone. Firstly, by encoding real-valued data vectors into compact binary codes, hashing makes efficient in-memory storage of massive data feasible in resource-limited user devices, and in an analogous way, the utilization of binary codes can reduce communication overheads as well. Secondly, as similarity calculation by inner product in a continuous vector space is replaced by bit operations in a discrete Hamming space, the time complexity of linear search is significantly reduced. Thirdly, by encoding the continuous real-valued vector into discrete binary codes, we prove that our proposed LightFR model is capable of effectively avoiding the leakage of user's sensitive information. Overall, the main contributions of this work are listed as follows:

\begin{itemize}
\item We tackle the problems of efficiency and privacy toward FRS in a unified way, \textit{i.e.}, the heavy parameterization inherited from real-valued representations in Euclidean space. In light of this, we seek solutions from the Hamming space by exploiting learning to hash technique, in which high-quality binary codes are obtained in the server and clients. To the best of our knowledge, this work represents the first effort towards this target in FRS.

\item To effectively train the discrete parameters in federated settings, we propose an efficient federated discrete optimization strategy between the central server and distributed clients, which facilitates both efficient and effective retrieval in terminal devices. Besides, we discuss the superiority of its multiple beyond-accuracy metrics, \textit{i.e.}, storage/communication efficiency, inference efficiency, and privacy preservation from the theoretical perspective.

\item Extensive experiments on four datasets with different volumes demonstrate the advantages of our model on effectiveness, efficiency and privacy over several state-of-the-art FRSs.

\end{itemize}

\section{Related Work}

In this section, we briefly review three relevant areas to this work, \textit{i.e.}, matrix factorization, learning to hash and federated recommender system. For a more comprehensive summary of the corresponding directions, please refer to the survey papers~\cite{cf_survey_2014,hash_survey_2017,frs_survey_2020}.

\subsection{Matrix Factorization}

Matrix factorization (MF), also known as latent factor model, has become a popular direction for collaborative filtering family in recommender systems~\cite{ecf_2004,mf_2009}. The goal of MF is to map original users and items into a common latent subspace, in which the similarities between users and items are calculated by inner products using their latent vectors~\cite{inner_product_2020}. Formally, assume that there are $n$ users, $m$ items and a user-item rating matrix $\mathbf{R}\in \mathbb{R}^{n\times m}$ in a website and the latent vector of user $u$ and item $i$ is denoted as $f$-dimensional embeddings, $\mathbf{p}_u\in \mathbb{R}^f$ and $\mathbf{q}_i\in \mathbb{R}^f$, so the observed rating $r_{ui}$ of user $u$ on item $i$ is estimated by the inner product of respective latent vectors, \textit{i.e.}, $\hat{r}_{ui}=\mathbf{p}_u^T\mathbf{q}_i$. A general objective is to minimize the following squared loss with regularization term:

\begin{equation}\label{eq:mf}
\begin{aligned}
{\min_{\mathbf{p}_u,\mathbf{q}_i}\sum_{(u, i, r_{ui}) \in \Omega}\left(r_{u i}-\mathbf{p}_{u}^{T} \mathbf{q}_{i}\right)^{2}+\lambda\left(\|\mathbf{P}\|_{F}^{2}+\|\mathbf{Q}\|_{F}^{2}\right) }
\end{aligned}
\end{equation}

\noindent where $\mathbf{P}\in\mathbb{R}^{f\times n}$ and $\mathbf{Q}\in\mathbb{R}^{f\times m}$ are the user and item latent matrix composed of all user and item latent vectors, respectively. Besides, $\Omega$ is a set of triplets of observed entries and $\lambda>0$ is a trade-off hyper parameter to avoid the over-fitting problem. The above loss function can be solved by (stochastic) gradient descent or alternating least square algorithms.

Owing to its high capability and flexibility, MF has attracted a lot of attention for many years. Early studies mainly focused on how to fuse side information via traditional mechanisms to improve recommendation performance~\cite{mf_context_2014,mf_social_2018}. Koren proposed SVD++ model by incorporating implicit feedback into MF method which only exploits explicit ratings~\cite{svd_plus_2008}. Hu et al. introduced the influence of geographical neighbors, business's review and category information into the delicate matrix factorization~\cite{mf_context_2014}, and again proved its high flexibility. Apart from fusing more information, MF can also be seamlessly integrated with other advanced models~\cite{mf_lda_2010,ncf_2017,mlp4rec_2022}. Agarwal et al proposed to introduce latent dirichlet allocation (LDA) into MF framework~\cite{mf_lda_2010}, where the use of an LDA prior is to regularize item factors and the combination of them can provide interpretable user factors as affinities to latent item topics. He et al. proposed neural collaborative filtering (NCF)~\cite{ncf_2017}, which incorporates a multi-layer perceptron into MF and can better model the user-item interactions with non-linear transformations. In short, a number of studies have investigated the superiority of fusing side information~\cite{cf_survey_2014} or complicated models~\cite{graph_survey2021} to enhance vanilla MF.

\subsection{Learning to Hash}

Recently, hashing has gained increasing attention due to its great efficiency in retrieving relevant items from massive data. The goal of hashing is to construct a mapping function to index each data point into a compact binary code, where the Hamming distances of similar
objects are minimized and that of dissimilar ones are maximized. There are two main kinds of hashing-based methods, \textit{i.e.}, locality sensitive hashing (LSH)~\cite{lsh_1998} and learning to hash (L2H)~\cite{hash_survey_2017}, where the formers are data-independent and use predefined hash functions without considering the underlying dataset, while the latters are data-dependent and learn tailored hash functions for specific datasets. Despite an extra training process, recent work showed that L2H greatly surpasses LSH in querying efficiency.

Studies in L2H have proceeded along two dimensions: two-stage approaches~\cite{col_hash_twostage_2014,bccf_2012} and learning hash codes directly~\cite{sdh_2015,dcf_2016}. For this research line of two-stage approaches, the first stage is to learn continuous representations for data, which are subsequently binarized into hashing codes using \textit{sign} threshold as a separate post-processing step. For instance, Zhou et al. learned user-item features with traditional CF and then rotated their learned features by running Iterative Quantization (ITQ) to acquire hash codes~\cite{bccf_2012}. However, such two-stage approaches are well-known to suffer from a large quantization loss, which is one of the main reasons why researchers are turning to the investigation of learning hash codes directly, where the binary codes are optimized straightforwardly rather than through a two-step approach. For example, Zhang et al. learned hash codes of users and items directly and further investigated additional constraints to improve generalization by better utilizing the Hamming space~\cite{dcf_2016}. Following that, Zhang et al. proposed a hashing based deep learning framework to unify the user-item interactions and the item content data to overcome the issues of data sparsity and cold-start, while improving the efficiency of online recommendation~\cite{ddl_2018}. However, the training process of these hashing methods mentioned above is usually conducted on centralized data. Hence, they are heavily not suitable for FRS scenario, where it has distinct advantages on privacy protection over centrally stored recommender systems, which is exactly the main motivation of our work.

\subsection{Federated Recommender System}

Federated Learning (FL) is a promising machine learning paradigm in recent years since it can enable collaborative learning across a variety of clients without sharing local private data~\cite{fl_first_2017,fl_survey_2019}. In general, there are two major components in the standard FL framework, where one is the client which trains the local models on their private user data independently, and the other is the server which aggregates the local models (gradients) uploaded from the clients to the global one. As a result of its role in ensuring privacy protection, there are many efforts to improve the basic FL 
framework, such as FedAvg~\cite{fl_first_2017}, FedProx~\cite{fedprox_2020} and FedRep~\cite{fedrep_2021}. In recommendation scenario, user private information, \textit{e.g.}, user's attribute and behavior interactions with items, is considerably sensitive information and probably cause identity information leakage if attacked by malicious parties~\cite{privacy_leakage_2008,wang2022trustworthy}. Hence, some recent endeavors have developed federated recommender system (FRS) for user privacy preservation while still maintaining considerable performance~\cite{frs_survey_2020,fmss_2022}. Federated Collaborative Filtering (FCF)~\cite{fcf_2019} and FedRec~\cite{fedrec_2020} are two pioneering \textit{privacy-by-design} works establishing a novel federated learning framework to learn the user and item embeddings on top of matrix factorization and the former is designed for implicit feedback, while the latter is for explicit feedback. However, Chai et al. argue that the model updates sent to the server in the original real-valued form as the aforementioned approaches do, may contain sensitive information to uncover raw data~\cite{fedmf_2020}. Along this path of research, Li et al. proposed to employ differential privacy to limit the exposure of the data in FRS~\cite{frs_dp_2020}. Besides, 
FedMF introduced homomorphic encryption into the FCF to ensure the confidentiality of parameter transmission~\cite{fedmf_2020}.

Aside from the privacy issue in FRS framework, there exists a great efficiency challenge on storing the global model in clients and transmitting the whole parameters between server and clients. From this research line, some efforts adopt meta learning mechanisms to reduce the payload on the clients~\cite{privrec_2021,metamf_2020}. For instance, Wang et al proposed to employ the approximated first-order gradients for one-stage meta learning, thereby reducing computational burden while maintaining a comparable performance~\cite{privrec_2021}. Besides, Lin et al. proposed MetaMF to deploy a big meta network into the server while deploying a small recommender model into the device to perform rating prediction~\cite{metamf_2020}. However, MetaMF may pose some privacy concerns since it necessitates the procedure of initializing the embeddings for all users on the server side. Additionally, there is also outstanding work that leverages knowledge distillation methodology to achieve the goal of deploying lightweight recommender models on user devices~\cite{fedkd_2022,llrec_2020}. Specifically, Wang et al. introduced LLRec framework, whose efficiency and robustness are maintained via the teacher-student training protocol, and then to perform the next point-of-interest recommendation task locally on resource-constrained clients~\cite{llrec_2020}. However, all these methods fail to take into account the challenges of efficiency (\textit{i.e.}, memory, calculation and communication) and privacy at the same time, which are the two major issues for real-world FRS. To better demonstrate the advantages of our approach, we summarize the qualitative comparisons between LightFR and existing FRS methods on efficiency and privacy in Table~\ref{tab:compfrs_analysis}. As for the quantitative analysis of the aforementioned aspects, please refer to Fig.~\ref{fig:non_acc_client} in Section 4.2.1 to see the related experimental results.

\begin{table}[t]
\renewcommand\arraystretch{1.2}
\footnotesize
  \caption{Comparison of different FRS methods relating to memory efficiency, inference efficiency, communication efficiency and privacy enhancement. The Eff. and Enh. denote Efficiency and Enhancement, respectively.}
\begin{center}
  \setlength\tabcolsep{5.5pt}
  \begin{tabular}{lcccc}
    \toprule
    \textbf{Models} & \textbf{Memory Eff.}  & \textbf{Inference Eff.}  & \textbf{Communication Eff.} & \textbf{Privacy Enh.}   \\
    \midrule
    \textbf{FCF}\cite{fcf_2019}          & $\times$  & $\times$ & $\times$ & $\times$ \\
    \textbf{FedMF}\cite{fedmf_2020}        & $\times$ & $\times$ & $\times$ & $\checkmark$ \\
    \textbf{FedRec}\cite{fedrec_2020}       & $\times$ & $\times$ & $\times$ & $\checkmark$ \\
    \textbf{MetaMF}\cite{metamf_2020}       & $\checkmark$ & $\times$ & $\checkmark$ & $\times$ \\
    \textbf{PrivRec}\cite{privrec_2021}       & $\times$ & $\times$ & $\checkmark$ & $\checkmark$ \\
    \textbf{LightFR}  &  $\checkmark$ & $\checkmark$ & $\checkmark$ & $\checkmark$ \\
    \bottomrule
  \end{tabular}
  \end{center}
  \label{tab:compfrs_analysis}
\end{table}

\begin{table}[h]
\renewcommand\arraystretch{1.2}
\footnotesize
  \caption{The list of notations and corresponding explanations.}
\begin{center}
  \setlength\tabcolsep{4.5pt}
  \begin{tabular}{cl}
    \toprule
    \textbf{Notation}  & \textbf{Explanation}\\
    \midrule
    $\mathcal{U}$; $\mathcal{I}$    & user (client) set; item set  \\
    $n$; $m$                     & total number of users; total number of items \\
    $\mathbf{R}$; $r_{ui}$                     & rating matrix whose element $r_{ui}$ denotes the rating score for user $u$ on item $i$ \\
    $u,v$; $i,j$                     & the specific user $u,v\in \mathcal{U}$; the specific item $i,j \in \mathcal{I}$  \\
    $\Omega_u; \Omega_i$ & local private dataset in client $u$; global feedback set interacted with item $i$ \\
    $\mathcal{I}_u;\mathcal{U}_i$ & the local observed item set of client $u$; global user set who interacts with item $i$ \\
    $\mathbf{p}_u; \mathbf{q}_i$ & the real-valued embedding of user $u$; real-valued embedding of item $i$ \\
    $\Delta \mathbf{q}_i^u$ & the gradient towards the item embedding vector $\mathbf{q}_i$ from the client $u$ \\
    $\mathbf{P}$; $\mathbf{Q}$ & the user real-valued embedding matrix; item real-valued embedding matrix \\
    $\mathbf{b}_u; \mathbf{d}_i$ & the binary vector of user $u$; binary vector of item $i$ \\
    $b_{uk}$; $b_{u\bar{k}}$ & the $k$-$\operatorname{th}$ bit code of the user binary vector $\mathbf{b}_u$; the rest codes of $\mathbf{b}_u$ excluding the $b_{uk}$ \\ 
    $\Delta d_{ik}^u$ & the gradient towards the $k$-$\operatorname{th}$ bit of item binary vector $\mathbf{d}_i$ from the client $u$ \\
    $\mathbf{B}$; $\mathbf{D}$ & the user binary embedding matrix; item binary embedding matrix \\
    $\Delta \mathbf{D}_t^u$ & the gradient matrix towards item binary matrix $\mathbf{D}$ uploaded from client $u$ at iteration $t$ \\
    $\mathcal{U}_s$ & a subset of clients randomly selected by the coordinated server\\
    $\eta$; $\lambda$ & the learning rate; regularization parameter\\
    $T$; $E$  & the number of training rounds between the server and clients; number of local training epochs \\
    $p$  & the fraction of clients to participate in current training round\\
    \bottomrule
  \end{tabular}
  \end{center}
  \label{tab:notation}
\end{table}

\section{The Proposed LightFR Framework}

In this section, we first formally describe the preliminaries, and then introduce the details of our proposed LightFR framework for efficient and privacy-preserving recommendation, followed by the federated discrete optimization algorithm delicately designed for FRS. Finally, we thoroughly discuss its superiority on multiple beyond-accuracy metrics from a
theoretical perspective. For clarity, we list some notations frequently used throughout the work in Table~\ref{tab:notation}.

\subsection{Preliminary}

Unlike fully centralized recommender systems, FRS hardly establishes a complete user-item rating matrix $\mathbf{R}^{n\times m}$ in the server since it no longer retains the fully observed dataset $\Omega$ for the sake of users' privacy. Specifically, we assume that there are a set of independent users $\mathcal{U}$, and a set of items $\mathcal{I}$ stored in the central server\footnote{The meaning of the terms ``user", ``client", and ``device" is same under federated settings, and we use them interchangeably.}. Following the FL principles, each user $u\in \mathcal{U}$ owns a local private dataset $\Omega_u$ consisting of some feedback tuples $(u,i,r_{ui} | i\in\mathcal{I}_u)$, where $\mathcal{I}_u$ represents observed items of client $u$. The goal of FRS is to predict the rating of user $u$ to each unseen item $i\in \mathcal{I}{\backslash}\mathcal{I}_u$ and then recommend the top ranked items to the target user. For federated training process, the gradients of users and items are calculated locally with the Eq.(\ref{eq:mf}). Specifically, the local update for their own user embedding $\mathbf{p}_u$ is performed independently without requiring any other user's private data:

\begin{equation}
\begin{aligned}
\mathbf{p}_u&=\mathbf{p}_u - 2\eta \left({\sum}_{i\in\mathcal{I}_u}(\mathbf{p}_u^T\mathbf{q}_i-r_{ui})\mathbf{q}_i + \lambda\mathbf{p}_u \right)
\end{aligned}
\end{equation}

\noindent where $\eta$ is the learning rate. Conversely, the item embedding $\mathbf{q}_i$ is updated globally on the server by aggregating local item gradient $\Delta \mathbf{q}_i^u$ uploaded from each client $u$:

\begin{equation}
\begin{aligned}
\mathbf{q}_i&=\mathbf{q}_i - 2\eta \left({\sum}_{u\in\mathcal{U}_i}\Delta \mathbf{q}_i^u \right)
\end{aligned}
\end{equation}

\noindent where the item gradient $\Delta \mathbf{q}_i^u=(\mathbf{p}_u^T\mathbf{q}_i-r_{ui})\mathbf{p}_u+\lambda \mathbf{q}_i$ is calculated by the local client $u$, and then the updated item matrix $\mathbf{Q}$ is sent down to all clients. The federated process repeats until convergence. For federated testing period, the client $u$ downloads the up-to-date item embedding matrix $\mathbf{Q}$ from FRS server and estimates ratings by inner product, \textit{i.e.}, $r_u=\mathbf{p}_u^T\mathbf{Q}$ in local device. 

\begin{figure}[t]
\begin{center}
\includegraphics[scale=0.47]{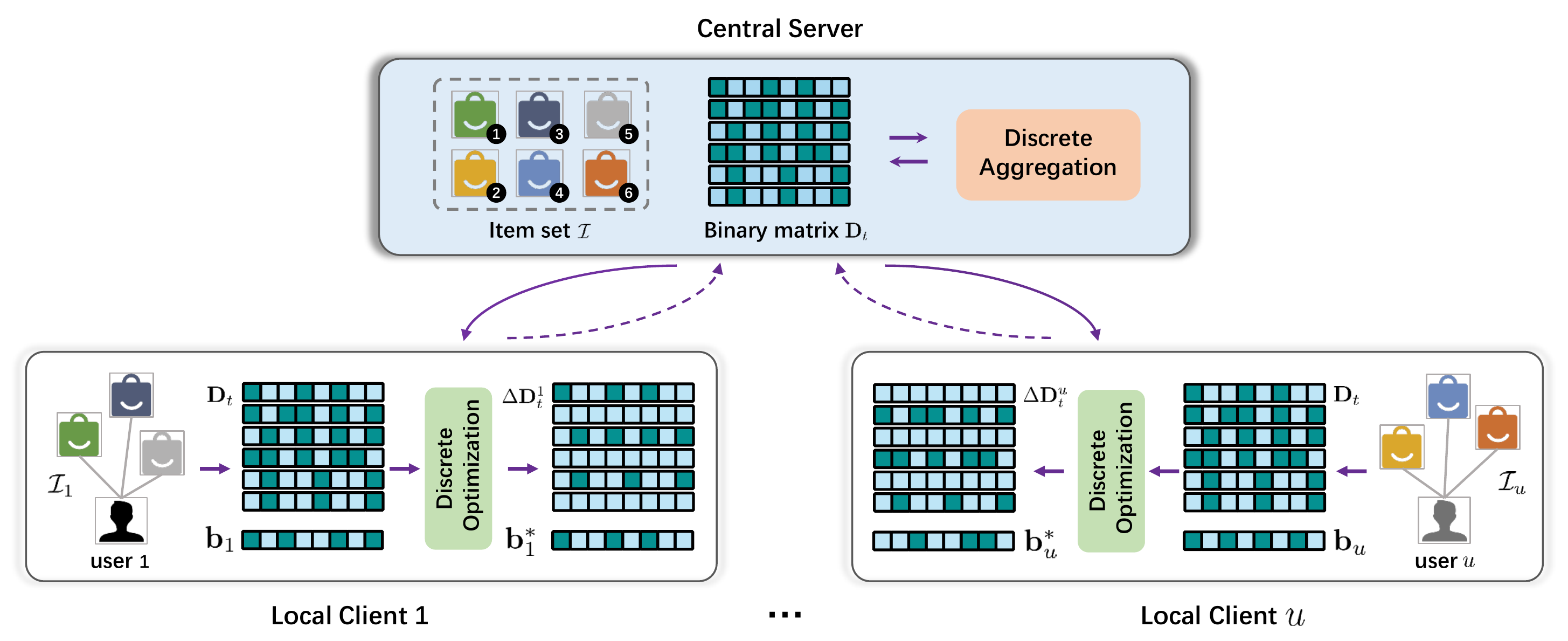}
\end{center}
\caption{The framework of our proposed LightFR approach. Firstly, the global item binary matrix $\mathbf{D}$ is initialized in the central server\protect\footnotemark, and then delivered to the distributed clients for local updates. Subsequently, each local client receives the latest item binary matrix $\mathbf{D}$ and (in the first round) initializes his own private binary vector $\mathbf{b}_u$. Later on, each client updates his own binary vector $\mathbf{b}_u$ and calculates the item gradient matrix $\Delta \mathbf{D}^u$ to be uploaded through the (local) discrete optimization module. After receiving the gradient matrix uploaded by all selected clients, the update process is conducted via the (global) discrete aggregation module on the server side. Finally, the latest item binary matrix is distributed to each client for next-step local optimization.}
\label{fig:lightfedrec}
\end{figure}
\footnotetext{For simplicity, we omit the subscript $t$ which denotes the $t$-$\operatorname{th}$ iteration in image caption part.}

\subsection{The LightFR Model}

As mentioned earlier, to fulfill the training process of FRS, the item embedding matrix $\mathbf{Q}$ and corresponding gradient matrix $\Delta \mathbf{Q}=[\Delta \mathbf{q}_1,\Delta\mathbf{q}_2,\cdots,\Delta\mathbf{q}_m]$ are exchanged between the server and clients. We emphasize that the parameters scale linearly in Euclidean space with the increasing number of items (as shown in Fig.~\ref{fig:stcomp}), posing a significant efficiency bottleneck in terms of storage, communication and inference time in resource-constrained local devices. Besides, transmitting gradient information in its raw 
real-valued form could result in privacy issues. In light of these challenges, we present our efficient and privacy-preserving method, with the assistance of binary codes generated by learning to hash technique in Hamming space, to the point where it is suitable for FRS when deployed in production.

An overview of the proposed framework is shown in Fig.~\ref{fig:lightfedrec}. In the beginning, the server randomly initializes an item binary matrix $\mathbf{D}=[\mathbf{d}_1,\cdots,\mathbf{d}_m]\in\{\pm 1\}^{f\times m}$ and clients initialize their own user binary vector $\mathbf{b}_u\in \{\pm 1\}^{f}$. Note that the server holds the entire item set $\mathcal{I}$, while each user $u$ exclusively has his/her private consumed item set $\mathcal{I}_u$. At iteration $t$, a subset of users $\mathcal{U}_s$ is randomly selected, and then each user $u\in \mathcal{U}_s$ downloads the latest global model (\textit{i.e.}, the item binary matrix $\mathbf{D}_t$) to the local device. Subsequently, the private user binary vector $\mathbf{b}_u$ is updated to $\mathbf{b}_u^*$ and the gradient of the item binary matrix $\Delta\mathbf{D}^u_t$ is calculated by (local) discrete optimization module using private local dataset $\Omega_u$. After the central server receives all local gradients submitted by users $\mathcal{U}_s$, it aggregates the collected gradients to facilitate the global model update by (global) discrete aggregation module. Finally, the server sends the latest item binary matrix to each client for the next round of optimization. The federated discrete optimization process is repeated until it converges. By jointly optimizing the pre-defined hashing-based loss functions between the server and clients, we can obtain the well-trained private user binary vectors in each client and 
the resulting item binary matrix in the server. Next we will introduce in detail the loss function of learning binary codes for users and items in FRS scenario.

The objective of LightFR is to tackle a joint discrete optimization task in federated settings, instead of solving the continuous optimization problem widely explored in traditional FRS. As a result, the similarities of two binary vectors cannot be measured directly by the inner product operations which will result in significant efficiency bottlenecks in the case of large-scale items, and hence the Hamming similarity is utilized to assess the proximity between the two ones. Instead of the Euclidean space, our proposed method aims to find binary codes in Hamming space, guaranteeing efficient storage and fast similarity search across users and items in federated scenarios. Generally, the similarity between user $u$ and item $i$ in Hamming space is defined as: 
\begin{equation}
\begin{aligned}
&\operatorname{sim}(\mathbf{b}_u, \mathbf{d}_i)=\frac{1}{f} \sum_{k=1}^{f} \mathbb{I}\left(b_{u k}=d_{i k}\right) \\
&=\frac{1}{2 f}\left(\sum_{k=1}^{f} \mathbb{I}\left(b_{u k}=d_{i k}\right)+f-\sum_{k=1}^{f} \mathbb{I}\left(b_{u k} \neq d_{i k}\right)\right) \\
&=\frac{1}{2 f}\left(f+\sum_{k=1}^{f} b_{u k} d_{i k}\right)\\
&=\frac{1}{2}+\frac{1}{2 f} \mathbf{b}_{u}^{T} \mathbf{d}_{i}
\end{aligned}
\end{equation}

\noindent where $b_{uk}$ and $d_{ik}$ denotes the $k$-$\operatorname{th}$ bit of the user and item binary codes $\mathbf{b}_u$ and $\mathbf{d}_i$, respectively. Besides, $\mathbb{I}(\cdot)$ represents the indicator function that yields 1 if the statement is true and 0 otherwise. Apparently, the value range of the Hamming similarity is ranging from 0 to 1, which just satisfies the basic requirements for similarity and $sim(\mathbf{b}_u, \mathbf{d}_i)=0$ if all the bits of $\mathbf{b}_u$ and $\mathbf{d}_i$ are totally different and $sim(\mathbf{b}_u, \mathbf{d}_i)=1$ if $\mathbf{b}_u=\mathbf{d}_i$. Note that the Hamming similarity has a highly efficient hardware-level implementation, which allows us to find similar items in time that is independent to the total number of items~\cite{binary_embedding_gpu_2018}. 

Similar to the problem of conventional MF in Eq.(\ref{eq:mf}), the preferences between users and items should be preserved by the above similarities with their respective binary codes, and the user-item rating matrix should be reconstructed by that as well. Therefore, the objective of our proposed LightFR built on $\operatorname{FedAvg}$~\cite{fl_first_2017} is formulated as follows:

\begin{equation}\label{eq:lightfedrec}
  \begin{aligned}
    \mathcal{L}=\sum_{u\in \mathcal{\mathcal{U}}}
    \frac{|\mathcal{I}_u|}{N}&
    \underbrace{\sum_{i \in \Omega_u}\left(r_{u i}-\operatorname{sim}\left(\mathbf{b}_{u}, \mathbf{d}_{i}\right)\right)^{2}}_{\text{local-specific loss}} + \underbrace{\lambda (||\sum_{u} \mathbf{b}_{u}||^{2}+||\sum_{i} \mathbf{d}_{i}||^{2})}_{\text{balanced constraint term}}\\
    & s.t. \   \mathbf{b}_u \in \{\pm1\}^{f},\mathbf{d}_i \in \{\pm1\}^{f}
  \end{aligned}
\end{equation}

\noindent where $N$ is the total number of instances over all clients, and $|\mathcal{I}_u|$ denotes the length of local samples on client $u$. Note that, due to the binary constraints above, the conventional regularization term $\|\mathbf{B}\|_{F}^{2}+\|\mathbf{D}\|_{F}^{2}$ used in Eq.(\ref{eq:mf}) becomes constant and hence is removed in Eq.(\ref{eq:lightfedrec}). However, in order to obtain the informative binary representations, a balanced constraint is utilized to maximize the information entropy of each bit, given in the form of constraint term in Eq.(\ref{eq:lightfedrec}), and the trade-off parameter $\lambda$ controls the proportion between minimizing the squared loss and the balanced constraints. Following that, we will detail our proposed federated discrete optimization method for optimizing the loss function Eq.(\ref{eq:lightfedrec}) in FRS scenario.

\subsection{Federated Discrete Optimization}

The goal of our federated discrete optimization is to find appropriate binary codes for users and items in federated settings such that the user preference over items is accurately preserved in Hamming space with their respective binary codes. However, solving the discrete optimization problem in Eq.(\ref{eq:lightfedrec}) by straightforward heuristics is challenging since it is generally an NP-hard problem which involves $\mathcal{O}(2^{(m+n)f})$ combinatorial searches for the binary codes. To this end, we introduce a collaborative alternating optimization method that can solve the above-mentioned question in a computationally tractable fashion in federated settings. Specifically, the proposed optimization algorithm mainly consists of two modules, \textit{i.e.}, (local) discrete optimization in each client and (global) discrete aggregation in the central server. Thereinto, the local discrete optimization module is primarily responsible for updating their respective user binary vectors and calculating the gradients of the binary item matrix to be uploaded using local data at the client; while the principal mission of the global discrete optimization module is to aggregate the gradients of items from multiple clients for updating the discrete item latent matrix. Next, we will elaborate on each module at length.

\subsubsection{(Local) Discrete Optimization}

In this part, we will introduce how to update the private user binary vector and calculate the gradients for item binary matrix with their private data leaving locally. We employ an alternating optimization strategy for solving the federated discrete optimization problem shown in Eq.(\ref{eq:lightfedrec}) that iterates the following two steps: (1) minimization with regard to each $\mathbf{b}_u$ with $\mathbf{d}_i$ fixed in clients; and (2) minimization with regard to $\mathbf{d}_i$ with $\mathbf{b}_u$ fixed by computing the gradients in clients and aggregating them in the server. Concretely, we in turn calculate the gradients for $\mathbf{b}_u$ and $\mathbf{d}_i$, given another one fixed, and then update the private user binary vector using the locally computed user gradients and upload the item binary gradients to the server for aggregation. 

First, we aim to optimize the private user binary vector $\mathbf{b}_u$ via fixing the item binary vector $\mathbf{d}_i$ in his/her own client without accessing any data from other clients. We update the user binary vector $\mathbf{b}_u$ of each client in parallel according to the following expanded formulation:

\begin{equation}\label{eq:lightfedrec_u}
\begin{gathered}
\underset{\mathbf{b}_{u} \in\{\pm 1\}^{f}}{\arg \min } \,\mathcal{L}^u_{local}= \frac{1}{4 f^{2}}\left(\sum_{i \in \Omega_{u}} \mathbf{d}_{i}^{T} \mathbf{b}_{u}\right)^{2}-\left(\sum_{i \in \Omega_{u}} \frac{2 r_{u i}-1}{2 f} \mathbf{d}_{i}^{T}\right) \mathbf{b}_{u} +\lambda\left\| \mathbf{b}_{u}\right\|^{2}
\end{gathered}
\end{equation}

\noindent where $\Omega_u$ denotes the private observed ratings for local client $u$ and the constant term containing $\mathbf{d}_i$ is omitted. We can easily verify that only the user's local data can be utilized to update the user's discrete representations so as to achieve the aim of privacy protection. Since the balanced constraint is plugged into the basic discrete loss function in Eq.(\ref{eq:lightfedrec}), the aforementioned minimization issue is generally NP-hard, and hence we employ the Discrete Coordinate Descent (DCD) algorithm~\cite{sdh_2015} to update each bit of the user binary vector $\mathbf{b}_u$. Specifically, let $b_{uk}$ denote the $k$-$\operatorname{th}$ bit of the user binary vector $\mathbf{b}_u$ and $\mathbf{b}_{u\bar{k}}$ be the rest binary codes excluding the $k$-$\operatorname{th}$ bit $b_{uk}$ of user $u$. Without loss of generality,  assume $\mathbf{b}_u=[b_{uk}\, \mathbf{b}_{u\bar{k}}]$ and $\mathbf{d}_i=[d_{ik}\, \mathbf{d}_{i\bar{k}}]$, and the quadratic term of Eq.(\ref{eq:lightfedrec_u}) with regard to $b_{uk}$ can be represented as:

\begin{equation}\label{eq:quadratic}
\begin{aligned}
&\frac{1}{4 f^{2}} \sum_{i \in \Omega_{u}}\left(\mathbf{d}_{i}^{T} \mathbf{b}_{u}\right)^{2}=\underbrace{\frac{1}{4 f^{2}} \sum_{i \in \Omega_{u}}\left(\left(\mathbf{d}_{i \bar{k}}^{T} \mathbf{b}_{u \bar{k}}\right)^{2}+\left(d_{i k} b_{u k}\right)^{2}\right)}_{\text {constant }}
&+\frac{1}{2 f^{2}} b_{u k} \sum_{i \in \Omega_{u}}\left(\mathbf{d}_{i \bar{k}}^{T} \mathbf{b}_{u \bar{k}} d_{i k}\right)
\end{aligned}
\end{equation}

\noindent where the constant part in Eq.(\ref{eq:quadratic}) can be omitted. Besides, the rest terms in Eq.(\ref{eq:lightfedrec_u}) with regard to $b_{uk}$ can be written as:

\begin{equation}\label{eq:restlfr}
\begin{aligned}
&\left(\sum_{i \in \Omega_{u}} \frac{2 r_{u i}-1}{2 f} \mathbf{d}_{i}^{T}\right) \mathbf{b}_{u}-\lambda\left\| \mathbf{b}_{u}\right\|^{2}\\
&=\underbrace{\sum_{i \in \Omega_{u}} \frac{2 r_{u i}-1}{2 f} \mathbf{d}_{i \bar{k}}^{T} \mathbf{b}_{u \bar{k}}+b_{u k}^{2}+\left(\sum_{k^{\prime}} b_{u k^{\prime}}\right)^{2}}_{\text {constant }}\\
&+b_{u k} \sum_{i \in \Omega_{u}} \frac{2 r_{u i}-1}{2 f} d_{i k}-2 \lambda b_{u k} \sum_{k^{\prime}} b_{u k^{\prime}}
\end{aligned}
\end{equation}

\noindent where $b_{u k^{\prime}}$ represents the $k^{\prime}$-$th$ bit of the $\mathbf{b}_{u\bar{k}}$. By bringing Eq.(\ref{eq:quadratic}) and Eq.(\ref{eq:restlfr}) into Eq.(\ref{eq:lightfedrec_u}) and omitting the constant parts, therefore, we can derive a series of bit-wise minimization problems:

\begin{equation}
\begin{aligned}
\underset{b_{u k} \in\{\pm 1\}}{\arg \min }\,\mathcal{L}^u_{local}&=\sum_{i \in \Omega_{u}} \frac{b_{uk}}{f}\left(\frac{1}{2 f} \mathbf{d}_{i \bar{k}}^{T} \mathbf{b}_{u \bar{k}}+\frac{1}{2}- r_{u i}\right) d_{i k} + 2 \lambda b_{uk} \sum_{k^{\prime}} b_{u k^{\prime}}\\
&=-b_{uk}\cdot \left[ \sum_{i \in \Omega_{u}} \frac{1}{f}\left(r_{u i}-\frac{1}{2}-\frac{1}{2 f} \mathbf{d}_{i \bar{k}}^{T} \mathbf{b}_{u \bar{k}}\right) d_{i k} -2 \lambda \sum_{k^{\prime}} b_{u k^{\prime}}\right]\\
&=-b_{uk}b_{uk}^{*}
\end{aligned}
\end{equation}

\noindent where $b_{uk}^{*}=\sum_{i \in \Omega_{u}} \frac{1}{f}\left(r_{u i}-\frac{1}{2}-\frac{1}{2 f} \mathbf{d}_{i \bar{k}}^{T} \mathbf{b}_{u \bar{k}}\right) d_{i k} -2 \lambda \sum_{k^{\prime}} b_{u k^{\prime}}$. We can clearly check that the optimized $b_{uk}$ should be the \textit{sign} operation of $b_{uk}^{*}$, according to the DCD update protocol where it will update $b_{uk}$ with $b_{u\bar{k}}$ fixed. Hence, the user binary vector $\mathbf{b}_u$ can be derived \textit{bit by bit} with the following update rule:

\begin{equation}\label{eq:user_update}
b_{u k}=\operatorname{sign}\left(F\left(b_{u k}^{*}, b_{u k}\right)\right)
\end{equation}

\noindent where $F(\cdot)$ is a custom function that $F(b_{u k}^{*},b_{u k})=b_{u k}^{*}$ if $b_{u k}^{*} \neq 0$ and $F(b_{u k}^{*},b_{u k})=b_{u k}$ otherwise. In other words, we do not update $b_{uk}$ when $b_{uk}^{*}=0$. In this way, the user binary vector $\mathbf{b}_u$ could be iteratively updated using their own local data until there is no change for each bit.

Secondly, we aim to calculate the gradients towards the item binary vector $\mathbf{d}_i$ via fixing the user binary vector $\mathbf{b}_u$, and then send it to the server for the preparation of global discrete aggregation. Note that different from the problem of updating user binary codes, the client $u$ only interacts with the item $i$ once in its local private data and can not access the feedback about item $i$ from other clients, hence we can update $\mathbf{d}_i$ according to the following expanded formulation:

\begin{equation}\label{eq:item_client}
\underset{\mathbf{d}_{i} \in\{\pm 1\}^{f}}{\arg \min } \,\mathcal{L}^i_{local}=\frac{1}{4 f^{2}}\left(  \mathbf{b}_{u}^{T} \mathbf{d}_{i}\right)^{2}-\left( \frac{2 r_{u i}-1}{2 f} \mathbf{d}_{i}^{T}\right) \mathbf{d}_{i}+\lambda\left\|\mathbf{d}_{i}\right\|^{2}
\end{equation}

\noindent where we only focus on the specific client $u$ in the process of local discrete optimization, and hence the sum operation (\textit{i.e.}, $\sum_u\in \Omega_i$) is prohibited in its local device. Finally, we can derive a series of bit-wise minimization problems on each bit of the item binary vector:

\begin{equation}\label{eq:item_grad}
\begin{aligned}
\underset{d_{i k} \in\{\pm 1\}}{\arg \min }\, \mathcal{L}_{l o c a l}^{i} &= \frac{d_{i k}}{f}\left(\frac{1}{2 f} \mathbf{b}_{u \bar{k}}^{T} \mathbf{d}_{i k} +\frac{1}{2}-r_{u i}\right) b_{u k}+2 \lambda d_{i k} \sum_{k^{\prime}} d_{i k^{\prime}} \\
&=-d_{i k} \cdot\left[\frac{1}{f}\left(r_{u i}-\frac{1}{2}-\frac{1}{2 f} \mathbf{b}_{u \bar{k}}^{T} \mathbf{d}_{i k}\right) b_{u k}-2 \lambda \sum_{k^{\prime}} d_{i k^{\prime}}\right] \\
&= -d_{ik}\cdot \left[\frac{1}{f}\Delta d^u_{ik} -2 \lambda \sum_{k^{\prime}} d_{i k^{\prime}} \right] \\
\end{aligned}
\end{equation}

\noindent where we define $\Delta d^u_{ik}=\left(r_{u i}-\frac{1}{2}-\frac{1}{2 f} \mathbf{b}_{u \bar{k}}^{T} \mathbf{d}_{i \bar{k}}\right) b_{u k}$ as the gradient of the $k$-$\operatorname{th}$ bit of the item binary vector $\mathbf{d}_i$ from the client $u$, and then upload it to the central server for aggregation. So far, the update of user binary vectors and the calculation of the gradients towards the item binary vectors have been completed in local discrete optimization module. Next, we will introduce the global discrete aggregation module to update the item binary vectors in the central server.

\subsubsection{(Global) Discrete aggregation} In this part, we will illustrate how to update the item binary matrix $\mathbf{D}$ using the gradients uploaded from the subset of clients $\mathcal{U}_s$ in the central server. Specifically, the loss function in the form of aggregation for the item binary vector $\mathbf{d}_i$ is as follows:

\begin{equation}\label{eq:item_server}
\underset{\mathbf{d}_{i} \in\{\pm 1\}^{f}}{\arg \min }\, \mathcal{L}^i_{global}= \frac{1}{4 f^{2}}\left(\sum_{u \in \Omega_{i}}  \mathbf{b}_{u}^{T} \mathbf{d}_{i}\right)^{2}-\left(\sum_{u \in \Omega_{i}} \frac{2 r_{u i}-1}{2 f} \mathbf{d}_{i}^{T}\right) \mathbf{d}_{i}+\lambda\left\|\mathbf{d}_{i}\right\|^{2}
\end{equation}

\noindent where $\Omega_i$ denotes the client set who has interacted with item $i$, which demonstrates that the same item will be associated with multiple clients. It's worth noting that this step involves data aggregation across different clients, so it needs to be executed on the server. Similar to the derivation process of updating the user binary vector, we can get a set of problems involving bit-wise minimization:

\begin{equation}\label{eq:item_global}
\begin{aligned}
\underset{d_{i k} \in\{\pm 1\}}{\arg \min }\, \mathcal{L}_{global}^{i} &=\sum_{u\in \Omega_i} \frac{d_{i k}}{f}\left(\frac{1}{2 f} \mathbf{b}_{u \bar{k}}^{T} \mathbf{d}_{i k} +\frac{1}{2}-r_{u i}\right) b_{u k}+2 \lambda d_{i k} \sum_{k^{\prime}} d_{i k^{\prime}} \\
&=-d_{i k} \cdot\left[\sum_{u\in \Omega_i}\frac{1}{f}\left(r_{u i}-\frac{1}{2}-\frac{1}{2 f} \mathbf{b}_{u \bar{k}}^{T} \mathbf{d}_{i k}\right) b_{u k}-2 \lambda \sum_{k^{\prime}} d_{i k^{\prime}}\right] \\
&=-d_{ik}\cdot\left[ \sum_{u\in \Omega_i}\frac{1}{f}\Delta d^u_{ik}-2 \lambda \sum_{k^{\prime}} d_{i k^{\prime}} \right]=-d_{ik}d_{ik}^{*} \\
\end{aligned}
\end{equation}

\noindent where $\Delta d_{ik}^u$ denotes the gradients towards the $k$-$\operatorname{th}$ bit of the item binary vector $\mathbf{d}_i$ uploaded from the client $u$. It can be observed from Eq.(\ref{eq:item_global}) that the update of global item binary vectors can be completely performed by the aggregation of the gradients uploaded from the clients. After aggregating the gradients $\Delta d_{ik}^u$ from the selected clients $\mathcal{U}_s$, the update of the item binary vector $\mathbf{d}_i$ can be performed \textit{bit by bit} with the following protocol:

\begin{equation}\label{eq:dik}
d_{i k}=\operatorname{sign}\left(F\left(d_{i k}^{*}, d_{i k}\right)\right)
\end{equation}

Following aggregation of the gradient data from all clients, the server conducts the sign operation on them, and finally obtains the updated item binary matrix $\mathbf{D}$, which will be distributed to the clients for the next round of optimization until convergence. To offer a holistic view of discrete aggregation and facilitate batch implementation, we define $\Delta \mathbf{D}^u$ as the gradient matrix from the client $u$ where $\Delta \mathbf{D}^u=[\Delta \mathbf{d}_1^u, \cdots, \Delta \mathbf{d}_m^u]$ and $\Delta \mathbf{d}_m^u=[\Delta d_{m1}^u,\cdots,\Delta d_{mk}^u]$. Hence, we can provide the matrix form of the aggregation rule based on the uploaded \textit{gradients} from the clients:

\begin{equation}\label{eq:agg_grad}
\operatorname{Agg}_{grad}: \mathbf{D}=\operatorname{sign}(\frac{1}{f} \sum_{u=1}^n \Delta \mathbf{D}^u-2 \lambda \mathbf{D}^{\prime})
\end{equation}

\noindent where $\mathbf{D}^{\prime}=[\mathbf{d}^{\prime}_1,\cdots,\mathbf{d}^{\prime}_m]$ and $\mathbf{d}^{\prime}_m=[d_{m1}^{\prime},\cdots,d_{mk}^{\prime}]$. Besides, the element $d_{mk}^{\prime}$ is defined as $d_{mk}^{\prime}=\sum_{k^{\prime}} d_{m k^{\prime}}$ and $d_{m k^{\prime}}$ represents the $k^{\prime}$-$th$ bit of the rest codes $\mathbf{d}_{m\bar{k}}$ exceeding the $d_{mk}$. We name this aggregation mechanism $\operatorname{Agg}_{grad}$. Apart from the aggregation of the gradients from clients, we can also directly aggregate the item binary matrix which has been locally updated on the client, and the aggregation mechanism based on the uploaded discrete \textit{parameters} is as follows:

\begin{equation}
\operatorname{Agg}_{para}: \mathbf{D}=\operatorname{sign}(\sum_{u=1}^n\mathbf{D}^u)
\end{equation}

\noindent where $\mathbf{D}^u$ denotes the item binary matrix which has been updated locally using the Eq.(\ref{eq:item_client}) in the client $u$, and then is uploaded to the server for aggregation. Similarly, we refer to this aggregation mechanism as $\operatorname{Agg}_{para}$. We will examine the two aggregation mechanisms in the experimental part. Note that the storage/communication efficiency and privacy can be improved by exchanging the binary matrix instead of the real-valued one. Next, we will logically present further details concerning these steps between the distributed clients and the coordinated server.

\begin{algorithm}[!t]
\SetKwProg{Fn}{Function}{}{end}
\caption{Federated Discrete Optimization Algorithm}
\label{alg:ours}
\LinesNumbered 
\KwIn{Total number of clients $n$; Total number of items $m$ \\
  \quad \quad \quad The code length $f$; The trade-off parameter $\lambda$; The number of selected clients $c$ \\
  \quad \quad \quad The global training rounds $T$; The local training epochs $E$ \\}
\KwOut{Global item binary matrix $\mathbf{D}$ at server; local user binary vector $\mathbf{b}_u$ in each client $u$}

\textbf{Server executes:} \\
\textbf{Initialization:} The item binary matrix $\mathbf{D}_t\in \mathbb{R}^{f\times m}$, where $t=0$; \\

\For{each round $t=1,\cdots,T$}{
    $\mathcal{U}_s \gets$ randomly select a subset of clients with the ratio of $p=c/n$; \\
    \For{each client $u \in \mathcal{U}_s$ in parallel}{
        $\Delta \mathbf{D}^u_t \gets$  $\operatorname{ClientUpdate(} \mathbf{D}_t$; $u$; $t\operatorname{)}$; \tcp*[f]{(Local) discrete optimization}
    }
    $\mathbf{D}_{t+1}\gets$ Eq.(\ref{eq:agg_grad}); \tcp*[f]{(Global) discrete aggregation}
}

\textbf{Client executes:} \\
\Fn{$\operatorname{ClientUpdate}(\mathbf{D}_t$; $u$; $t \operatorname{):}$ }{
    downloading latest $\mathbf{D}_t$ from the central server\;
    \For{each epoch $e=1,\cdots,E$}{
        \For(\tcp*[f]{Update private user binary vector}){$k=1,\cdots,f$}{ 
            $b_{u k} \gets$ Eq.(\ref{eq:user_update});  \\
        }
        
        \For(\tcp*[f]{Computing gradients for item binary matirx}){$(i,r_{ui})\in \Omega_u$}{
            \For{$k=1,\cdots,f$}{
                $\Delta d^u_{ik}\gets \left(r_{u i}-\frac{1}{2}-\frac{1}{2 f} \mathbf{b}_{u \bar{k}}^{T} \mathbf{d}_{i \bar{k}}\right) b_{u k}$; \\
                $\Delta \mathbf{D}_{t}[k;i]\gets \Delta d^u_{ik}$; \\
            }
        }
    }
        
    return $\Delta \mathbf{D}_{t}$; \\
    
}
\end{algorithm}

The pseudo-code of the algorithm for Federated Discrete Optimization is presented in Algorithm 1. The input consists of the number of clients and items, $n$ and $m$ respectively, and the training hyper-parameters such as the code length $f$, global training rounds $T$ and local training epochs $E$. The target is to output the well-trained global item binary matrix in the central server and the private user binary vector in each client. In the algorithm, the Line 3 to the Line 9 is the loop operated on the server, which sends parameters to clients and collects their gradients for aggregation. The function $\operatorname{ClientUpdate()}$ is the operation on local devices where the Line 14 to Line 16 is the procedure for updating private user binary vector and the Line 17 to Line 22 is to calculate the gradients towards item binary matrix. Finally, this function returns the gradients (Line 24) to the server for global aggregation. The above training process, \textit{i.e.}, the outer loop (Line 3 to Line 9) will repeat until its convergence, \textit{e.g.}, the achievement of the preset training rounds or predefined thresholds.

Through the analysis of the above algorithm, we can conclude that its total time complexity is $\mathcal{O}(T\times n \times E\times (f^2\times |\Omega_u|+ m\times f^2))$. Note that, the optimization between clients can be easily performed in parallel under federated settings, so the number of clients $n$ here can be omitted. Therefore, the final time complexity of our proposed algorithm can be expressed as $\mathcal{O}(T \times E\times f^2 \times (|\Omega_u|+ m))$. Because $T$, $E$ and $f$ are usually small hyper-parameters in our work and fixed during the training stage, we can see that the training time complexity (\textit{a.k.a} the encoding time for the binary codes) is linear with the number of rated items per client (\textit{i.e.}, $|\Omega_u|$) and the number of items (\textit{i.e.}, $m$). In summary, training our discrete optimization algorithm is efficient under federated scenarios.

\subsubsection{Cold-start Scenario}

Necessarily, the cold-start issue, in which few or even no prior interactions (\textit{e.g.}, ratings or clicks) are known for certain users or items, is an inherited challenging problem in traditional collaborative filtering paradigms. Similarly,  it is also required to account for the existence of new clients (\textit{a.k.a.} cold-start clients) or new items (\textit{a.k.a.} cold-start items) in federated recommendation scenarios. Hence, in this part, we will introduce how to deal with the situation of new clients (items) in our federated discrete optimization algorithm.

Apparently, it is expensive to train the whole algorithm from scratch to obtain binary codes for these cold-start samples, when new users (clients) or items arrive. Therefore, a feasible countermeasure is to learn temporary binary codes for new coming samples online and then retrain the entire data offline when possible. Note that we focus on the cold-start situation which allows for the existence of a few interactions with users or items. As for the scenario where there is entirely no interactions, it must necessitate the assistance of some side information and warm-up techniques~\cite{meta_coldrs_2021,ucold_2022}, which is beyond the research scope of our work.

Firstly, we will explore the case when a new client arrives. Without loss of generality, let $\{ r_{ui} | i \in \Omega_u \}$ be the set of local observed private interactions for existing items in the new client $u$ and its binary codes is $\mathbf{b}_u$. It is worth mentioning that it's unnecessary to impose the global balanced constraint as described in Eq.(\ref{eq:lightfedrec}) for a single user. Hence, we should only concentrate on minimizing the squared error loss in each client in the following way:

\begin{equation}\label{eq:u_cold}
\begin{aligned}
\underset{\mathbf{b}_{u} \in\{\pm 1\}}{\arg \min }\,\mathcal{L}^u_{cold} = \sum_{i\in \Omega_u} \left(r_{u i}-\operatorname{sim}\left(\mathbf{b}_{u}, \mathbf{d}_{i}\right)\right)^{2}
\end{aligned}
\end{equation}

\noindent where $\operatorname{sim}(\mathbf{b}_u,\mathbf{d}_i)=\frac{1}{2}+\frac{1}{2 f} \mathbf{b}_{u}^{T}\mathbf{d}_i$. We can easily observe that Eq.(\ref{eq:u_cold}) is a particular form of Eq.(\ref{eq:lightfedrec_u}) with removing the regularization term. So we can quickly learn the $k$-$\operatorname{th}$ bit $b_{uk}$ of $\mathbf{b}_u$ by the DCD optimization protocol $b_{u k}=\operatorname{sign}\left(F\left(b_{u k}^{*}, b_{u k}\right)\right)$, and $b_{uk}^{*}$ is derived from the following formulation:

\begin{equation}\label{eq:u_cold_update}
\begin{aligned}
b_{uk}^{*} = \sum_{i \in \Omega_{u}} \frac{1}{f}\left(r_{u i}-\frac{1}{2}-\frac{1}{2 f} \mathbf{d}_{i \bar{k}}^{T} \mathbf{b}_{u \bar{k}}\right) d_{i k}
\end{aligned}
\end{equation}

\noindent where $\mathbf{b}_{u\bar{k}}$ denotes the rest codes of the user binary vectors $\mathbf{b}_u$ excluding the $k$-$\operatorname{th}$ bit $b_{uk}$. We can see that for the arrival of new users, we can update the user binary vectors locally via the discrete optimization module in their terminal devices without retraining a large number of existing user binary vectors.

Secondly, we will explore the case when a new item arrives. Similar to the procedure of cold-start clients, the global balanced constraint specified in Eq.(\ref{eq:item_server}) is ignored for a new coming item, and the following loss function in the server can be modified as:

\begin{equation}\label{eq:i_cold}
\begin{aligned}
\underset{\mathbf{d}_{i} \in\{\pm 1\}}{\arg \min }\,\mathcal{L}^i_{cold} = \sum_{u\in \Omega_i} \left(r_{u i}-\operatorname{sim}\left(\mathbf{b}_{u}, \mathbf{d}_{i}\right)\right)^{2}
\end{aligned}
\end{equation}

\noindent where $\Omega_i$ denotes the set of global observed interactions for existing clients on target item $i$, which indicates that the new coming item will be interacted across multiple clients. Hence, we will perform the gradient aggregation process by the discrete aggregation module in the server and conduct the gradient calculation procedure via the discrete optimization module on each client. By expanding and simplifying the above Eq.(\ref{eq:i_cold}), we can acquire the aggregation form on the server as follows:

\begin{equation}\label{eq:i_cold_agg}
\begin{aligned}
\underset{\mathbf{d}_{i} \in\{\pm 1\}}{\arg \min }\, \mathcal{L}_{cold}^{i} &=\sum_{u\in \Omega_i} \frac{d_{i k}}{f}\left(\frac{1}{2 f} \mathbf{b}_{u \bar{k}}^{T} \mathbf{d}_{i k} +\frac{1}{2}-r_{u i}\right) b_{u k}\\
&=-d_{i k} \cdot\sum_{u\in \Omega_i}\frac{1}{f}\left(r_{u i}-\frac{1}{2}-\frac{1}{2 f} \mathbf{b}_{u \bar{k}}^{T} \mathbf{d}_{i k}\right) b_{u k} \\
&=-d_{ik}\cdot\left[ \sum_{u\in \Omega_i}\frac{1}{f}\Delta d^u_{ik} \right]=-d_{ik}d_{ik}^{*} \\
\end{aligned}
\end{equation}

\noindent where $d_{ik}^{*}$ denotes the gradient aggregation process in the server and $\Delta d_{ik}^u$ denotes the gradient calculation procedure towards the $k$-$\operatorname{th}$ bit of the item binary vector $\mathbf{d}_i$ uploaded from the client $u$. By performing the gradient calculation  to the new items on each client and uploading them to the server for aggregation, the near-optimal update of the binary codes towards those cold-start items is achieved on the premise of protecting users' local privacy information.

\subsection{Discussion}

In this section, we theoretically discuss the superiority of our proposed $\operatorname{LightFR}$ from three beyond-accuracy perspectives: storage/communication efficiency, inference efficiency, and privacy preserving.

\subsubsection{Storage/Communication Efficiency}

In federated settings, the storage overhead in local clients and communication consumption between the server and clients are always an unneglectable issue~\cite{client_reduction_2018}. As for the storage overhead in each client, it is mainly composed of the private user embedding and the global item embedding matrix. Considering the traditional Euclidean space which is widely applied in many FRS methods and a 64-bit floating point precision, the simple formulation of storage overhead estimation is exactly: $((1+m)\times f \times 64)/8$ bytes, which increases linearly with the ever-increasing number of items. For example, we assume the number of items $m$ is 10 million and the dimension $f$ is 128, and it will take over 10.2 GB of storage space, which is hard to be deployed into general devices with limited memory. Apparently, by storing the binary representations of the user and items in Hamming space, the memory consumption will not exceed 1.3 GB, which is acceptable for common mobile devices. As for the communication overhead which is exchanged between the server and users, it primarily depends on the number of items $m$ to recommend. The requirement to transmit huge parameters (\textit{i.e.}, the item embedding matrix) between the FL server and users over several communication rounds imposes strict limitations for both the server and clients. In Euclidean space, the formula used to estimate communication overhead is $(m\times f\times 64)/8$ while the approximated formula in Hamming space is only $m\times f$. Similar to the calculation process of storage overhead, the communication consumption in Euclidean space is 8 times than that in Hamming space. To be more precise, the summarized formulas in Table~\ref{tab:non_acc_formulas} and experimental results in Fig.~\ref{fig:non_acc_client} comparing with other classical federated recommender systems in storage and communication aspects clearly show the efficiency of our method. As a result, by transmitting the binary item matrix produced by our LightFR model, the payload of data exchanged can be considerably reduced, and thus allowing user devices to utilize lower bandwidth resources.

\subsubsection{Inference Efficiency}

Unlike many existing FRS approaches, which estimate the correlation scores via inner product or cosine similarity in a continuous Euclidean space between user and item representations, our proposed LightFR model eventually in each client generates their own private binary vectors and acquires the latest global item binary matrix which is downloaded from the server, and then efficiently perform the similarity search in Hamming space using the binary ones at their local devices. Generally, given $f$-dimensional representation of $m$ items in Euclidean space and the results of top-$k$ recommendations will incur an inference inefficiency with the time complexity of $\mathcal{O}(mf+k\operatorname{log}k)$, which scales approximately linearly with the number of items. Not surprisingly, our proposed method adopts bit operations in a proper Hamming space, so the time complexity of linear search is 
greatly decreased and even constant time scan is possible~\cite{spectral_hashing_08}. Besides, the Hamming similarity has a highly efficient hardware-level implementation, allowing us to locate relevant items in time that is independent to the number of items~\cite{binary_embedding_gpu_2018}. To be specific, if items are represented by the $f$-dimensional double-precision float embeddings, and the inner product of them in Euclidean space requires $f$ times of floating-point multiplications, while the similarity calculation with $f$-dimensional binary vectors in Hamming space needs only one XOR operation and one time of sum operation. The experimental results quantified in the third panel in Fig.~\ref{fig:non_acc_client} verify the efficiency of our method in inference time compared with other federated recommender systems. In a nutshell, the similarity search in Hamming space is more significantly efficient than that in Euclidean space, and it is more urgent and suitable for resource-constrained clients in the FRS scenarios.

\subsubsection{Privacy Preserving}

It is critical to preserve and enhance privacy in FRS scenarios, since previous work has proved that the original rating data is likely to be leaked when transmitting the gradients or model parameters in real-valued forms~\cite{privacy_leakage_2008}. The proposed LightFR requires the exchange of binary representations in discrete Hamming space rather than the real-valued ones in continuous Euclidean space between the server and clients, which is the key property that brings benefits to FRS in terms of privacy enhancement. As FedMF~\cite{fedmf_2020} states, given the real-valued gradients of a user $u$ towards the item $i$ at iterations $t$ and $t+1$ uploaded in two continuous steps \textit{i.e.}, $g^t_{i}$ and $g^{t+1}_{i}$, and the corresponding real-valued user latent embedding $\mathbf{p}_u$, it can infer the user's rating information $r_{ui}$ according to the following formulations:

\begin{align}
\frac{g_{i k}^{t}}{p_{u k}^{t}}-\frac{g_{i k}^{t+1}}{p_{u k}^{t}+\frac{\alpha_{k}}{p_{u k}^{t}}}&=\frac{p_{u k}^{t}}{g_{i k}^{t}} \beta_{i}+\frac{g_{i k}^{t}}{p_{u k}^{t}} \gamma_{i} \label{eq:gradient_leak1}\\
r_{u i}=\frac{g_{i k}^{t}}{p_{u k}^{t}}&+\sum_{d=1}^{f} p_{u d}^{t} q_{i d}^{t} \label{eq:gradient_leak2}
\end{align}

\noindent where $g_{ik}^t$ denotes the $k$-$\operatorname{th}$ dimension of the uploaded gradient $g_i^t$ about the item $i$ at iteration $t$, and $p_{ud}^t$ and $q_{id}^t$ represent the $d$-$\operatorname{th}$ dimension of the user embedding $\mathbf{p}_u$ and item embedding $\mathbf{q}_i$, respectively. Besides, we can treat $\beta_i$, $\gamma_i$ and $\alpha_k$ as the constants. Hence, the premise of inferring the rating information $r_{ui}$, \textit{i.e.}, the Eq.(\ref{eq:gradient_leak2}), is to solve the variable $p_{uk}^t$. We can easily confirm that there must be one real-valued scalar of $p_{uk}^t$ in Euclidean space that satisfies the Eq.(\ref{eq:gradient_leak1}), and it can be solved by some iterative methods to compute a numeric solution, \textit{e.g.}, gradient descent optimization methods. 
Notably, our proposed method assumes that the user's latent embedding $p_{uk}^t$ is discrete in Hamming space, which violates the premise of the continuous real-valued solutions in Eq.(\ref{eq:gradient_leak2}). Besides, the non-differentiable and discontinuous operation \textit{sign}, which is an irreversible process, makes it tough or even impossible to solve the Eq.(\ref{eq:gradient_leak1}). Thus, the users' private rating data on their local devices can not be easily inferred. 

In addition, we provide a theoretical analysis to demonstrate that our proposed LightFR model is able to enhance users' privacy. Assuming $d_{ik}^{t+1}$ is the $k\operatorname{-}$th bit of item $i$ to be uploaded from user $u$ to the server at time $t+1$, according to Eq.(\ref{eq:dik}), we have
\begin{align}
d_{ik}^{t+1} = sign(F((d_{ik}^*)^t, d_{ik}^t)) \nonumber\\
(d_{ik}^*)^t = \frac{1}{f}(r_{ui} - \frac{1}{2} - \frac{1}{2f}(d_{i\bar{k}}^T)^t b_{u\bar{k}}^t)b_{uk}^t \label{eq:dik*t}
\end{align}
where $F(x, y)$is a function that $F(x, y) = x$ if $x \neq 0$ and $F(x, y) = y$ otherwise. Thus, we will discuss the effectiveness of preserving users' privacy according to the following three cases.
\begin{enumerate}
    \item if $d_{ik}^{t+1} \neq d_{ik}^t$ and $(d_{ik}^*)^t \neq 0$, we can determine whether $(d_{ik}^*)^t$ is positive or negative by $d_{ik}^{t+1}$. When we get sign($(d_{ik}^*)^t$) and  $(d_{i\bar{k}}^T)^t$, we have no idea to determine $r_{ui}$ from Eq.(\ref{eq:dik*t}) since $\mathbf{b}_u^t$ is unkonwn.
    \item if $d_{ik}^{t+1} = d_{ik}^t$ and $(d_{ik}^*)^t \neq 0$, similar with (1), we can only get sign($(d_{ik}^*)^t$). So the value of $r_{ui}$ is also undetermined.
    \item if $d_{ik}^{t+1} = d_{ik}^t$ and $(d_{ik}^*)^t = 0$, we have 
    \begin{align}\label{eq:dik_privacy}
        (d_{ik}^*)^t &= \frac{1}{f}(r_{ui} - \frac{1}{2} - \frac{1}{2f}(d_{i\bar{k}}^T)^t b_{u\bar{k}}^t)b_{uk}^t \\
       & =\frac{1}{f}(r_{ui} - \frac{1}{2} - \frac{1}{2f}((\mathbf{d}_i^T)^t\mathbf{b}_u^t)b_{uk}^t + \frac{1}{2f^2}d_{ik}^t = 0 \label{eq:dikeq0_privacy}
    \end{align}

    Following that, we expand the Eq.(\ref{eq:dik_privacy}) and Eq.(\ref{eq:dikeq0_privacy}) by the dimension $f$, and we can derive $r_{ui}$ from the following two sets of equations: 
    \begin{equation}\label{eq:t}
\left\{
             \begin{array}{lr}
             \frac{1}{f}(r_{ui} - \frac{1}{2} - \frac{1}{2f}((\mathbf{d}_i^T)^t\mathbf{b}_u^t)b_{u1}^t + \frac{1}{2f^2}d_{i1}^t = 0, &  \\
             \vdots \\
             \frac{1}{f}(r_{ui} - \frac{1}{2} - \frac{1}{2f}((\mathbf{d}_i^T)^t\mathbf{b}_u^t)b_{uk}^t + \frac{1}{2f^2}d_{ik}^t = 0, &   \\
             \vdots \\
             \frac{1}{f}(r_{ui} - \frac{1}{2} - \frac{1}{2f}((\mathbf{d}_i^T)^t\mathbf{b}_u^t)b_{uf}^t + \frac{1}{2f^2}d_{if}^t = 0
             \end{array}
\right.
\end{equation}
\begin{equation}\label{eq:t+1}
\left\{
             \begin{array}{lr}
             \frac{1}{f}(r_{ui} - \frac{1}{2} - \frac{1}{2f}((\mathbf{d}_i^T)^{t+1}\mathbf{b}_u^{t+1})b_{u1}^{t+1} + \frac{1}{2f^2}d_{i1}^{t+1} = 0, &  \\
             \vdots \\
             \frac{1}{f}(r_{ui} - \frac{1}{2} - \frac{1}{2f}((\mathbf{d}_i^T)^{t+1}\mathbf{b}_u^{t+1})b_{uk}^{t+1} + \frac{1}{2f^2}d_{ik}^{t+1} = 0, &   \\
             \vdots \\
             \frac{1}{f}(r_{ui} - \frac{1}{2} - \frac{1}{2f}((\mathbf{d}_i^T)^{t+1}\mathbf{b}_u^{t+1})b_{uf}^{t+1} + \frac{1}{2f^2}d_{if}^{t+1} = 0
             \end{array}
\right.
\end{equation}

It is worth noting that the premise of Eq.(\ref{eq:t}) and Eq.(\ref{eq:t+1}) hold on is that $(d_{ik}^*)^t = 0, (d_{ik}^*)^{t+1} = 0, \forall k \in \{1,2,\cdots, f\}$. Such a premise illustrates the invalidity of the classical discrete coordinate descend optimization algorithm~\cite{sdh_2015}. Thus, when $d_{ik}^{t+1} = d_{ik}^t$ and $(d_{ik}^*)^t = 0$, the server cannot derive the value of $r_{ui}$. In summary, we cannot infer the sensitive rating data of client $u$ based on the uploaded information.
\end{enumerate}

Therefore, to some extent, our LightFR framework can theoretically prevent malicious attackers from inferring the sensitive rating information of local clients, thereby achieving the purpose of enhancing the capacity of preserving privacy.

\section{Experiments}

In this section, we first introduce our experimental settings in detail, and then present the extensive experimental results and in-depth analysis that validate the effectiveness of our proposed LightFR framework from multiple aspects.

\subsection{Experimental Settings}

First, we introduce the details of the adopted datasets in our work, and then elaborate on the evaluation metrics utilized to verify the effectiveness and efficiency of our proposed model. Besides, we list several comparison recommendation methods based on centralized storage and some privacy-preserving ones based on federated learning, and finally, we detail some other implementation details to ensure reproducibility and fair comparison of the experiments.

\begin{table}[h]
\renewcommand\arraystretch{1.2} 
\setlength\tabcolsep{20pt} 
\footnotesize
  \caption{Statistics of the utilized datasets in evaluation.}
\begin{center}
  \setlength\tabcolsep{4.5pt}
  \begin{tabular}{lrrrccc}
    \toprule
     \textbf{Datasets}  & \textbf{\# Users}  & \textbf{\# Items} & \textbf{\# Ratings} & \textbf{\# Average} & \textbf{Rating Range} & \textbf{Data Density}  \\
    \midrule
    MovieLens-1M~\cite{movielens_2015}  & $6,040$ & $3,952$ & $1,000,209$ & $166$ & $[1,2,\cdots,5]$ & $4.19\%$ \\
    Filmtrust~\cite{filmtrust_2013}     & $1,508$ & $2,071$ & $35,497$ & $24$ & $[0.5,1,\cdots,4]$  & $1.14\%$ \\
    Douban-Movie~\cite{douban_2011}      & $2,964$ & $39,695$ & $894,888$ & $302$ & $[1,2,\cdots,5]$ & $0.76\%$ \\
    Ciao~\cite{ciao_2013}    & $7,375$ & $105,096$ & $282,619$ & $38$ & $[1,2,\cdots,5]$ & $0.04\%$ \\
    \bottomrule
  \end{tabular}
  \end{center}
  \label{tab:datasets}
\end{table}

\subsubsection{Datasets}

For a comprehensive comparison, we adopt four commonly used public datasets with various scales to conduct experimental analyses, which are MovieLens-1M\footnote{https://grouplens.org/datasets/movielens/1M/}, Filmtrust\footnote{https://guoguibing.github.io/librec/datasets.html}, Douban-Movie\footnote{https://www.cse.cuhk.edu.hk/irwin.king.new/pub/data/douban} and Ciao\footnote{https://www.cse.msu.edu/~tangjili/datasetcode/truststudy.htm}. Specifically, MovieLens-1M dataset originally contains 
approximately 1 million ratings of 3,952 movies from 6,040 users, and the data density is 4.19\% and the average number of user ratings is 166. Filmtrust dataset is crawled from online rating website FilmTrust, which originally contains about 35 thousand ratings, from 1,508 users on 2,071 films and its data density is 1.14\% and each user has 24 ratings in average, which has the least interactions among them. Douban-Movie dataset is built from online sharing website Douban which provides user rating, review and recommendation services for movies, books and music, and it has nearly 894 thousand ratings from 2,964 users of 39,695 movies and its data density is only 0.76\% and the average number of ratings per user is 302 which is the most interactions among them. Ciao dataset is crawled from the popular product review sites Ciao in the month of May, 2011, and it contains about 282 thousand ratings on 105,096 movies from 7,375 users, which is the most sparse dataset with only the density ratio of 0.04\% and the average number of user ratings is merely 38. The rating scale of MovieLens-1M, Douban-Movie and Ciao ranges from 1 to 5 in 1 increment, while Filmtrust ranges from 0.5 to 4 in 0.5 increments. For each user, we first sort the positive samples by timestamp in chronological order and then separate them into three chunks: 80\% as the training set, 10\% as the validation set and 10\% as the test set. For the validation and testing stage in federated settings, we randomly sample a fixed number of items as negatives for each positive item in each local client. The detailed statistics of these datasets are summarized in Table~\ref{tab:datasets}, where \# Average means the average number of user ratings, which reflects the data density of clients in each dataset. Importantly, experiments conducted on the above four datasets with varying scales and sparsity can comprehensively reflect the performance of the model. In federated settings, each user is regarded as a local client, and the user’s data is locally stored on the device.

\subsubsection{Evaluation Metrics}

To evaluate the performance and verify the effectiveness of our model, we utilize two commonly used evaluation metrics, \textit{i.e.}, Hit Ratio (HR) and Normalized Discounted Cumulative Gain (NDCG), and both of them are widely adopted for item ranking task. The above two metrics are usually truncated at a particular rank level (\textit{e.g.} the first $k$ ranked items) to emphasize the importance of the first retrieved items. Specifically, HR@k, recall at a cutoff $k$, is used to count the number of occurrences of the testing item in the predicted ranked item set, and NDCG@k, which truncates the ranked list at $k$, measures the ranking quality which assigns higher scores to hit at the top position ranks while the positive items at bottom positions of the ranking list contribute less to the final result. Intuitively, the HR metric measures whether the test item is present on the top-$k$ ranked list or not, and the NDCG 
metric measures the ranking quality which comprehensively considers  both the positions of ratings and the ranking precision.

\subsubsection{Comparison Models}

We adopt two kinds of benchmarks for comprehensive comparisons, \textit{i.e.,} classical Matrix Factorization (MF) based methods and recent federated MF approaches. The classical MF-based methods are based on centralized storage settings, which are not capable of protecting user privacy. Most existing federated MF methods essentially perform similarity search via inner product in Euclidean space, which are not able to efficiently handle the rating data in a privacy enhancement manner. 

\textbf{Classical MF-based models}

\begin{itemize}
    \item PMF~\cite{pmf_2007}: a canonical probabilistic latent factor model which factorizes both users and items into a common subspace, in which the similarity between users and items can be measured by inner product in Euclidean space.
    \item SVD++~\cite{svd_plus_2008}: another latent factor model which explores the biases of users and items, and incorporates the user implicit feedback into PMF framework.
    \item DDL~\cite{ddl_2018}: a hashing-based MF model which adds deep learning technique into the discrete collaborative filtering framework by exploiting rating and item content data.
     \item NCF~\cite{ncf_2017}: the state-of-the-art deep learning based MF method that combines generalized matrix factorization and multi-layer perceptron (MLP) to model user-item interactions.
\end{itemize}

\textbf{Federated MF models}

\begin{itemize}
    \item FCF~\cite{fcf_2019}: a pioneering privacy-preserving federated collaborative filtering method which formulates the updating rules of collaborative filtering to suit the FL settings.
    \item FedMF~\cite{fedmf_2020}: a privacy-enhanced matrix factorization approach based on secure homomorphic encryption under federated settings.
    \item FedRec~\cite{fedrec_2020}: it is another privacy-enhanced model with non-cryptographic techniques, in which some unrated items are randomly sampled and assigned with some virtual ratings.
    \item MetaMF~\cite{metamf_2020}: a novel federated MF model that deploys a big meta network into the server while deploying a small model into the device to perform rating prediction task.
    \item PrivRec~\cite{privrec_2021}: a fast-adapting federated recommender model that adopts a meta-learning strategy to enable fast convergence on local devices in a privacy-preserving way\footnote{For fair comparison, we instantiate PrivRec equipped with the first-order meta learning and differential privacy components, which is built on the MF backbone as our proposed model does.}.
\end{itemize}

\subsubsection{Implementation Details}

In our experiments, the dimension of user and item embedding $f$ is set to 32 for all the real-valued MF methods and 64 for the hashing-based models. The significance of setting the dimensions in this way is to achieve the recommendation performance comparable to the real-valued models as much as possible on the premise of saving resources compared with those dense models. Therefore, the threshold of our LightFR being effective lies in the length of binary codes. For the centralized MF-based models, we set the training epoch $E$ to 50 and adopt the early stopping technique, that is, if the performance of five consecutive epochs is not improved, it will stop running, and set the batch-size $B$ to 512. For the federated MF-based comparison methods, we set the global training rounds $T$ to 50, the local epoch $E$ to 1. Besides, we set the ratio of selected clients $p$ to 0.6. Moreover, we specify the length of public key $l=1024$ used in FedMF~\cite{fedmf_2020}, the sampling parameter $\rho=3$ in FedRec~\cite{fedrec_2020}, the length of hidden layers $L=2$, the number of hidden units $h=8$ and the size of low-dimensional item embeddings $s=8$ in MetaMF~\cite{metamf_2020}, and the noise scale $z=1$, the clipping bound $S=50$ in PrivRec~\cite{privrec_2021}. All the hyper parameters are searched and tuned according to the performance on the validation dataset.

\subsection{Experimental Results}

In this section, we present the extensive experimental results of LightFR \textit{w.r.t.} state-of-the-art centralized and federated baselines. In detail, firstly we will compare our approach with the other nine recent MF-based methods in overall recommendation performance, and then the ablation study is provided to analyze the contribution of our proposed method. Finally, the sensitivity analysis is further given to explore the effects of different hyperparameters on our model.

\subsubsection{Overall Performance}

We conduct the overall comparison of different models including classical MF-based methods and federated MF models, where the first four methods are centralized manner and the last five methods are in federated settings. Table~\ref{tab:compfrs} summarizes the experimental results of HR@10 and NDCG@10 on the four widely used datasets. Firstly, we deliver the analysis of the classical MF-based models under the centralized storage paradigm. From such results, we have the following observations.

\begin{itemize}

\item Among the centralized classical MF models, the fact that SVD++ model outperforms PMF shows the advantage of incorporating implicit data into the explicit feedback on top of the basic MF framework. Besides, the reason why DDL is nearly comparable to PMF is that the deep neural model with item content data is introduced on the basis of discrete collaborative filtering algorithm, which is beneficial to extract efficient binary representations.

\item In addition, thanks to the effective non-linear feature transformation and high-order feature extraction capability of multi-layer perceptron (MLP), NCF surpasses the other two canonical MF models. Note that such improvement increases along with the increasing of data scale, where the datasets are arranged in the order of increasing data scale, which demonstrates that the deep neural models represented by MLP require a big quantity of data to work, which is resource-intensive to run on the client in the federated learning environment.

\end{itemize}

\begin{table}[h]
\renewcommand\arraystretch{1.4}
\footnotesize
  \caption{Comparison results of LightFR and the baselines on the four datasets. The best federated learning results are in bold and the second best federated method is with asterisks, and the best results for centralized learning methods are underlined.}
\begin{center}
  \setlength\tabcolsep{3.5pt}
  \begin{tabular}{lcccccccc}
    \toprule
     \multirow{2}{*}{\textbf{Methods}} &  \multicolumn{2}{c}{\textbf{MovieLens-1M}}  & \multicolumn{2}{c}{\textbf{Filmtrust}}  & \multicolumn{2}{c}{\textbf{Douban-Movie}} & \multicolumn{2}{c}{\textbf{Ciao}}   \\ \cline{2-9}
          & \textbf{HR@10}& \textbf{NDCG@10}& \textbf{HR@10}& \textbf{NDCG@10}& \textbf{HR@10}& \textbf{NDCG@10}& \textbf{HR@10}& \textbf{NDCG@10}   \\
    \midrule
     \textbf{PMF}\cite{pmf_2007}  &  0.5124  & 0.2768  & 0.8704 & 0.6610  & 0.3011 & 0.1678 & 0.4636  & 0.2434  \\
     \textbf{SVD++}\cite{svd_plus_2008}   &   0.5291  & 0.2826  & 0.8793  & 0.6777 & 0.3118 & 0.1886  & 0.4692  & 0.2463  \\
      \textbf{DDL}\cite{ddl_2018}   &   0.5101  & 0.2743  & 0.8630  & 0.6579 & 0.2967 & 0.1671 & 0.4566  & 0.2429  \\
     \textbf{NCF}\cite{ncf_2017}    & \underline{0.5342}  & \underline{0.2901} & \underline{0.8827} & \underline{0.6907}  & \underline{0.3223} & \underline{0.2097} & \underline{0.4732} & \underline{0.2513} \\ 
    \hline
    \textbf{FCF}\cite{fcf_2019}       & 0.4945  & 0.2625  & 0.8543  & 0.6376  & 0.2921  & 0.1615  & 0.4492 & 0.2304   \\
    \textbf{FedMF}\cite{fedmf_2020}    & 0.4836    & 0.2534  & 0.8601  & 0.6563   & 0.2786 & 0.1443  & 0.4461 & 0.2272 \\
    \textbf{FedRec}\cite{fedrec_2020}   &  0.4893  & 0.2612  & 0.8503   & 0.6304  & 0.2832 & 0.1593  & 0.4474 & 0.2301  \\
    \textbf{MetaMF}\cite{metamf_2020}    &  0.4994$^{*}$  & 0.2691$^{*}$  & 0.8566  & 0.6389  & \textbf{0.2937} & \textbf{0.1669}  & 0.4503 & 0.2402   \\
    \textbf{PrivRec}\cite{privrec_2021}    &  0.4989  & 0.2688  & 0.8605$^{*}$  & 0.6563$^{*}$  & 0.2931 & 0.1666  & 0.4534$^{*}$  & 0.2411$^{*}$   \\
    \hline
    \textbf{LightFR}  & \textbf{0.5014}  & \textbf{0.2709}  &  \textbf{0.8615} & \textbf{0.6565}  & 0.2934$^{*}$  & 0.1665$^{*}$   &\textbf{0.4556}  &\textbf{0.2413}  \\
    \bottomrule
  \end{tabular}
  \end{center}
  \label{tab:compfrs}
\end{table}

Next, we focus on the experimental analysis of the federated MF baselines and our proposed model. From the experimental results in Table~\ref{tab:compfrs}, the following noteworthy findings are drawn.

\begin{itemize}

\item The five federated MF baselines (such as FCF, FedMF, FedRec, MetaMF and PrivRec), marginally impair the performance compared with the centralized MF models (such as PMF, SVD++, DDL and NCF). More specifically, the recommendation performance of the FedMF and FedRec methods is often inferior to that of the centralized MF models, and the performance of the FCF, MetaMF and PrivRec models are comparable to that of the centralized ones. This is mainly because, in the federated settings, in order to achieve privacy protection and prevent the original data from leaving the local clients, the global parameters are updated by aggregating the gradient information or model parameters of the clients, and thus the gradient noises and model losses are inevitably introduced during the aggregation process.

\item Considering that the performance of FCF method is significantly better than that of the FedMF and FedRec methods. There are two reasons: on the one hand, the loss function of FCF is modeled based on implicit feedback data, whereas the FedMF and FedRec directly model the explicit feedback data. On the other hand, FedMF can be seen as introducing a homomorphic encryption mechanism on the basis of FCF, which often results in a small performance drop but strengthens the privacy protection ability of the FRS framework, while FedRec can be seen as introducing a hybrid filling strategy based on FCF, which means that it well address the privacy issue but introduce some noise to the raw data.

\item Among the five federated MF baselines, PrivRec model is superior to other baselines in most cases since it introduces a first-order meta-learning method that enables fast convergence and few communication rounds with only a few data points in local devices, which is particularly evident in Filmtrust and Ciao datasets. Besides, the performance of MetaMF is optimal on Douban-Movie dataset, since it introduces a meta network on top of collaborative filtering methods to capture the collaborative information, and its performance improves gradually with the rise of data volume, which shows that the sufficient training data is the cornerstone of high performance in deep neural models but it is not realistic on the local clients in federated scenario. It should be noted that the performance of MetaMF in Filmtrust dataset is worse than that of FedMF, which could be owing to the over-fitting issue caused by deep neural models in the case of little amounts of training data. Although these methods can achieve comparable performance, the transfer of the original model parameters between the server and clients may make it subject to privacy leakage attacks.

\item Our proposed model LightFR outperforms the federated baselines in most cases. Specifically, LightFR is superior to the five federated MF baselines (\textit{i.e.}, FCF, FedMF, FedRec, MetaMF and PrivRec) in terms of every metric on MovieLens-1M, Filmtrust and Ciao datasets, and the performance on Douban-Movie dataset is comparable to MetaMF. Moreover, the performance of our method is comparable to that of the centralized MF models such as PMF, and our model can be strengthened when more complex modeling techniques are introduced (such as SVD++ model with implicit feedback information and NCF model with high-order extracted features). Although our method is slightly worse than MetaMF method on Douban-Movie dataset, our method can achieve the purpose of less inference time, less memory occupation and fewer bandwidth resources under the premise of considerable performance, which can be easily migrated and deployed to mobile terminal devices under the federated settings, while MetaMF still performs nearest neighbor search via inner products with the real-valued embeddings in Euclidean space, which will result in significant storage consumption and calculation overhead.

\end{itemize}

It's worth noting that the recommendation performance of our proposed LightFR model can be on par with that of the real-valued models in most cases. The intriguing experimental results can be supported by the dimension of the user (item) latent vectors $f$. As mentioned above in the experimental settings, we set the dimension of the latent embeddings $f=32$ in real-valued MF models and $f=64$ for the binary MF models. Note that, our main motivation is to explore the lightweight and privacy enhancement mechanisms in federated recommendation scenario, so that the resource-constrained clients can occupy less memory and communication overheads, and shorten the inference time under the strict data privacy protections. Therefore, the purpose of this setting is to narrow the gap in recommendation accuracy between the binary codes and that of the real-valued ones on the premise of saving resources. In order to make a fair comparison, we still maintain the same experimental settings to measure the storage, communication overheads and inference time between the real-valued models and the binary models in the next part. Through extensive experiments, we show that our binary model can greatly reduce the storage, communication overheads and inference time while keeping the recommendation performance comparable. In future work, we will consider further improving the accuracy of our binary model when it uses the same code length as the real-valued models, which is beyond the scope of this current work.

\begin{table}[t]
\renewcommand\arraystretch{1.6}
\footnotesize
  \caption{Summary of formulas for estimating the space complexity, storage and communication cost.}
\begin{center}
  \setlength\tabcolsep{6pt}
  \begin{tabular}{lccc}
    \toprule
     \textbf{Methods}& \textbf{Space Complexity} &\textbf{Storage Cost (Bytes)}  & \textbf{Communication Cost (Bytes)}    \\
    \midrule
    \textbf{FCF}\cite{fcf_2019}          & $\mathcal{O}(|\mathcal{I}_u|+m)$   & $\left[\left(1+m\right)\times f \times 64\right]/8$  & $[(|\mathcal{I}_u|+m)\times f \times 64]/8$  \\ 
    \textbf{FedMF}\cite{fedmf_2020}        &  $\mathcal{O}(|\mathcal{I}_u|+m)$   & $\left[\left(1+m\right)\times f \times 1024\right]/8$ & $[(|\mathcal{I}_u|+m)\times f \times 1024]/8$   \\ 
    \textbf{FedRec}\cite{fedrec_2020}       & $\mathcal{O}(|\mathcal{I}_u|(1+\rho)+m)$    & $\left[\left(1+m\right)\times f \times 64\right]/8$ & $\{[|\mathcal{I}_u|(1+\rho)+m]\times f \times 64\}/8$  \\ 
    \textbf{MetaMF}\cite{metamf_2020}       &  $\mathcal{O}(m+L\cdot h)$   & $\left[\left(m+L\cdot h \right)\times f \times 64\right]/8$ & $[(m\times f +2\cdot L\cdot h +f\times s+ s\times m)\times 64]/8$  \\
    \textbf{PrivRec}\cite{privrec_2021}       & $\mathcal{O}(|\mathcal{I}_u|+m)$    & $\left[\left(1+m\right)\times f \times 64\right]/8$  & $[(|\mathcal{I}_u|+m)\times f \times 64]/8$ \\ 
    \textbf{LightFR}  & $\mathcal{O}(|\mathcal{I}_u|+m)$    & $\left(1+m\right)\times f$ & $ (|\mathcal{I}_u|+m)\times f$  \\
    \bottomrule
  \end{tabular}
  \end{center}
  \label{tab:non_acc_formulas}
\end{table}

\begin{figure}[t]
\begin{center}
\includegraphics[scale=0.5]{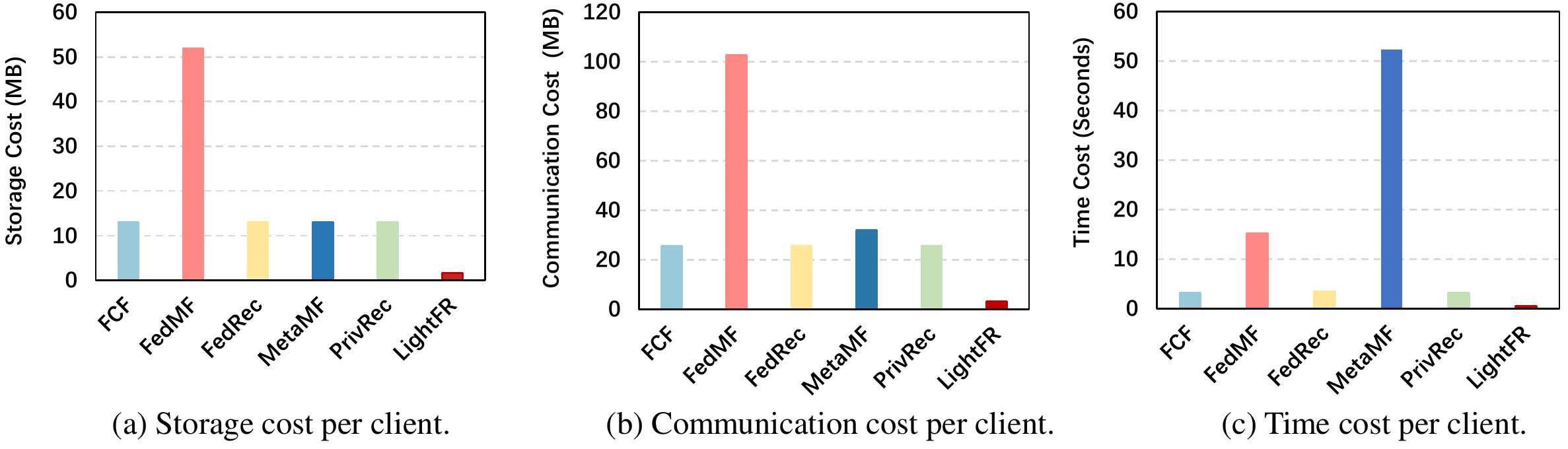}
\end{center}
\caption{Comparisons with baselines in terms of storage cost, communication cost and inference time cost on each local client.
}
\label{fig:non_acc_client}
\end{figure}

In this part, we compare our model with five federated MF baselines on several beyond-accuracy metrics, \textit{i.e.}, storage (memory) cost, communication cost and inference time cost on the local client, so as to verify the comprehensive performance of our method. Specifically, we take the large-scale Ciao dataset as an example, and calculate the storage overhead, communication consumption which includes uploads and downloads, and test the inference time cost on the client side. As for the calculation of memory and communication overheads, we identify and set the default parameters as follows: the number of items in Ciao dataset $m=105,096$, the average number of items interacted by user in Ciao dataset $\mathcal{I}_u=38$. Table~\ref{tab:non_acc_formulas} summarizes the formula list of the comparison methods (FCF, FedMF, FedRec, MetaMF and PrivRec) and our LightFR model to estimate the model space complexity, storage and communication overheads. And the first two panels of the Fig.~\ref{fig:non_acc_client} show the experimental results in terms of storage and communication costs on the client side. When it comes to the cost of inference time, we perform the similarity search on 
a local client with 16GB of RAM and 2.30GHz 8-core processor. The third panel of Fig.~\ref{fig:non_acc_client} demonstrates the inference time cost of the five federated MF baselines and our proposed method.

From the results of Fig.~\ref{fig:non_acc_client}, we can draw the following conclusions. First, our LightFR consistently outperforms the state-of-the-art federated MF baselines in terms of memory, communication and inference efficiency. More specifically, as for storage efficiency, the occupancy of LightFR is about 3.1\% of that in FedMF method, about 8.1\% of that in MetaMF model, and exactly 8\% of that in FCF, FedRec and PrivRec models in Fig.~\ref{fig:non_acc_client} (a). This is mainly because FedMF needs an encryption process, which expands the dimension of the original vector from 32 to 256; MetaMF needs to store the additional parameters of private Rating Prediction (RP) module, in addition to the item embeddings; And yet, FCF, FedRec, PrivRec and LightFR just need to retain the item embedding matrix, whereas the first three methods require the real-valued forms and our method only requires the binary ones. When it comes to communication efficiency in Fig.~\ref{fig:non_acc_client} (b), the advantage of LightFR is similar to the tread in storage efficiency. The LightFR only requires to transmit the lightweight binary item embeddings between the server and clients, while FCF needs to transmit the encrypted item embeddings which are larger than the original ones, and FCF, FedRec, MetaMF and PrivRec require to transfer the real-valued item embeddings or model parameters. In terms of search efficiency, LightFR incurs more significant speedup than the federated MF baselines, which is about 100 times faster than MetaMF, around 30 times faster than FedMF, and 7 times faster than FCF, FedRec and PrivRec models in Fig.~\ref{fig:non_acc_client} (c). Since MetaMF requires a forward propagation process to generate the predicted ratings for target user on some items, which is the main reason for its slow inference speed. Besides, FedMF must execute the further decryption process for the encrypted item embedding matrix before performing the similar search for all the existing items on the local client. Moreover, our LightFR is faster than FCF, FedRec and PrivRec, since the former performs the XOR operation employing the binary codes in Hamming space, whereas the latter two approaches utilize the real-valued embeddings to execute the inner product operations in Euclidean space. As a result, our LightFR 
model outperforms the other state-of-the-art federated baselines in terms of memory cost, communication overhead and inference time, while leads to negligible accuracy degradation.

\subsubsection{Ablation Study}

To better understand the contribution of our proposed federated discrete optimization algorithm, we evaluate the performance gain of our method over several variants in Table~\ref{tab:ablation}. We denote the full model as LightFR which performs fine-grained gradient aggregation process in federated discrete aggregation module. \textbf{LightFR}$_{para}$ represents the direct discrete parameter aggregation mechanism. \textbf{LightFR}$_{init}$ obtains binary codes by directly conducting median quantization on real-valued features learned by MF without the collaborative optimization procedure between the server and clients. \textbf{Random} means that the binary codes of the users and items are randomly generated at the clients and the similarity search is conducted in the Hamming space.

\begin{table}[h]
\renewcommand\arraystretch{1.4}
\footnotesize
  \caption{Ablation study results towards HR@10 and NDCG@10 on the four datasets.}
\begin{center}
  \setlength\tabcolsep{3.5pt}
  \begin{tabular}{|l|cc|cc|cc|cc|}
    \hline
     \multirow{2}{*}{\textbf{Methods}} &  \multicolumn{2}{c|}{\textbf{MovieLens-1M}} & \multicolumn{2}{c|}{\textbf{Filmtrust}}  & \multicolumn{2}{c|}{\textbf{Douban-Movie}} & \multicolumn{2}{c|}{\textbf{Ciao}}   
     \\ \cline{2-9}
          & \textbf{HR@10}& \textbf{NDCG@10}& \textbf{HR@10}& \textbf{NDCG@10}& \textbf{HR@10}& \textbf{NDCG@10}& \textbf{HR@10}& \textbf{NDCG@10}   \\
    \hline
    \textbf{Random}  & 0.2419  & 0.1043  &  0.5793 & 0.3531 & 0.1408  & 0.0893   & 0.2882  & 0.1254  \\
    \hline
     \textbf{LightFR}$_{init}$  & 0.3110  & 0.1542  &  0.6129 & 0.4932  & 0.1832  & 0.1193  & 0.3105  & 0.1632  \\
    \hline
      \textbf{LightFR}$_{para}$  & 0.4942  & 0.2618
  &  0.8589 & 0.6554  & 0.2913  & 0.1619   & 0.4489  & 0.2267  \\
    \hline
    \textbf{LightFR}  & \textbf{0.5014}  & \textbf{0.2709}  &  \textbf{0.8615} & \textbf{0.6565}  & \textbf{0.2934}  & \textbf{0.1665}   &\textbf{0.4556}  &\textbf{0.2413}  \\
    \hline
  \end{tabular}
  \end{center}
  \label{tab:ablation}
\end{table}

From the experimental results, we can draw some important conclusions. Firstly, the fact that our method is clearly superior to the \textbf{Random} method, which confirms its supremacy of our proposed discrete optimization algorithm in the federated settings. Besides, we find that the method \textbf{LightFR}$_{init}$ of discretizing the latent features of the pre-trained MF model is also less effective than our LightFR method, which verifies the superiority of our proposed method in the collaborative optimization between the server and local clients. Finally, the LightFR model, which performs gradient aggregation process on the server, is marginally better than the \textbf{LightFR}$_{para}$ which directly adopts parameter aggregation mechanism in the discrete aggregation module. The results can be ascribed to the fact that the use of direct aggregation of discrete parameters causes more information loss than the well-designed gradient aggregation mechanism in our approach.

\subsubsection{Sensitivity Analysis} In this part, we analyze the performance fluctuations of our proposed LightFR with varying hyper parameters including the length of binary codes $f$, the trade-off parameter $\lambda$, and the ratio of selected clients $p$. 

\begin{figure}[t]
\begin{center}
\includegraphics[scale=0.44]{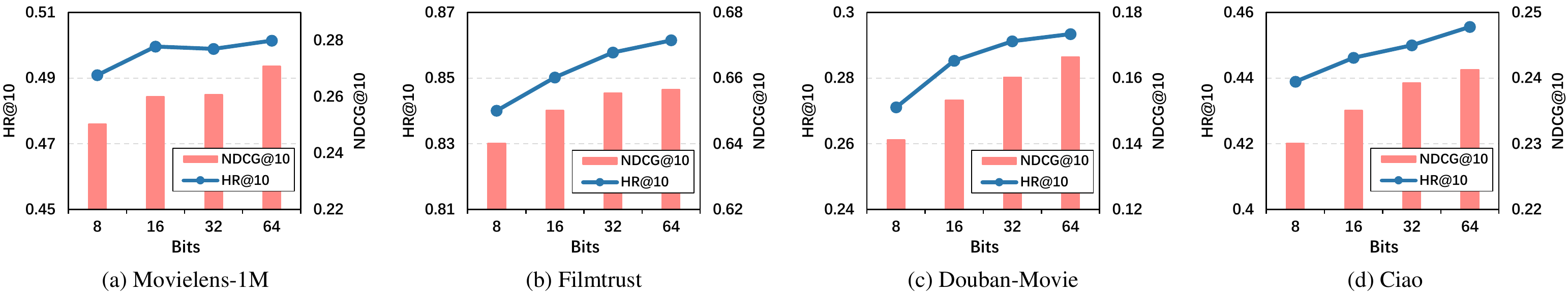}
\end{center}
\caption{Performance of LightFR with various code lengths $f$ evaluated on four datasets.
}
\label{fig:bit_length_lightfedrec}
\end{figure}

Firstly, the impact of the length of binary codes $f$ on performance is studied. According to the experimental results shown in Fig.~\ref{fig:bit_length_lightfedrec}, as $f$ increases from 8 to 64, significant performance improvements of LightFR are observed on the four datasets. Although with the increase of dimensions, the growth rate of performance gradually slows down. According to the above experimental results, the following insights can be obtained, that is, in the case of restricted storage capacity on the local client, the length of binary representations of users and items can be large as much as feasible, which can fully represent the structural properties of the original data.

\begin{figure}[t]
\begin{center}
\includegraphics[scale=0.44]{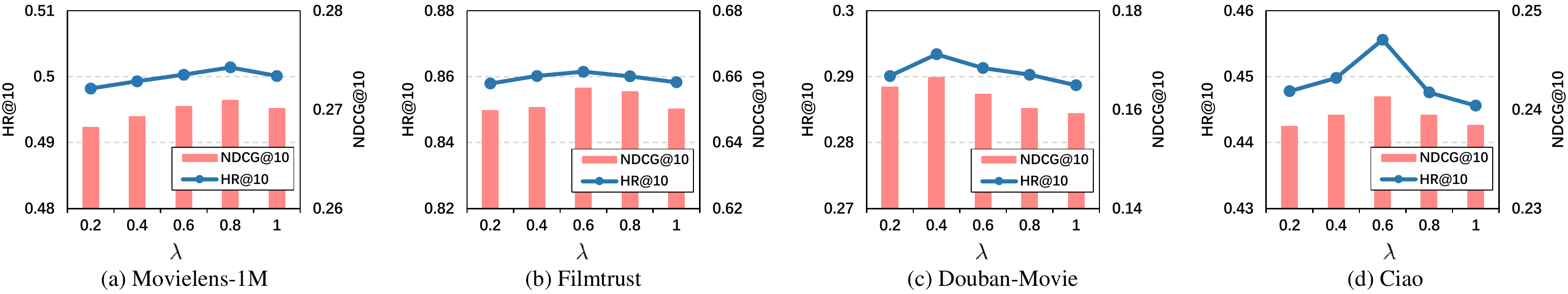}
\end{center}
\caption{Performance of LightFR with different values of trade-off parameter $\lambda$ evaluated on four datasets.
}
\label{fig:lambda_lightfedrec}
\end{figure}

Then, we experiment on a series of different values of trade-off parameter $\lambda \in \{0.2, 0.4, 0.6, 0.8 , 1\}$. As shown in Fig.~\ref{fig:lambda_lightfedrec}, when the value of the trade-off parameter $\lambda$ increases, the HR@10 and NDCG@10 of the model show a upward trend, indicating that the balanced constraints are of positive significance to the discrete modeling of user preferences and item attributes. However, when the value of the trade-off parameter continues to increase, the model performance does not show a corresponding improvement, indicating that too large the value of the trade-off parameter is not conducive to the binary representations of the users and items but may lead to the appearance of overfitting issue. As a result, with the increase of the trade-off parameter $\lambda$, the recommendation accuracy increases at first. Then, the accuracy will gradually decrease as the $\lambda$ continues to increase further. Besides, the change of the trade-off parameter $\lambda$ has little fluctuations in the overall performance, which shows that our method is somewhat insensitive to the hyper parameter of $\lambda$.

\begin{figure}[t]
\begin{center}
\includegraphics[scale=0.44]{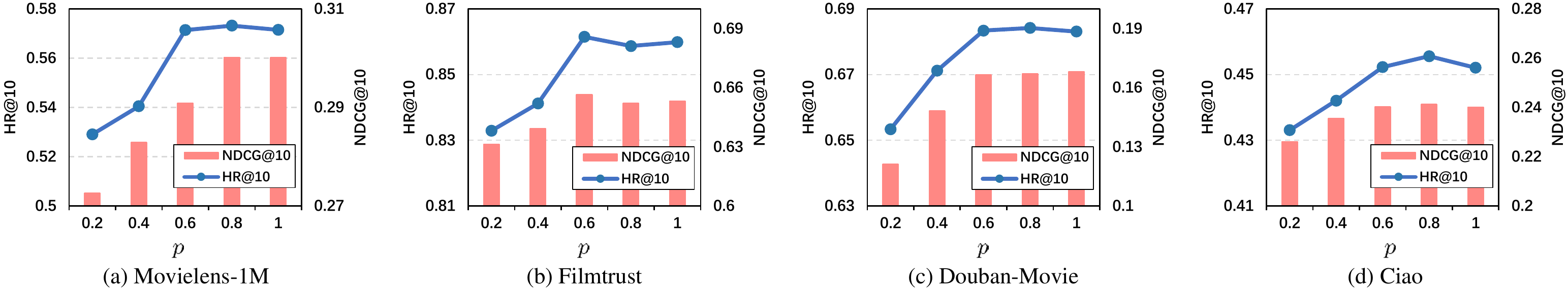}
\end{center}
\caption{Performance of LightFR with different ratios of selected clients $p$ evaluated on four datasets.
}
\label{fig:selected_ratio_lightfedrec}
\end{figure}

Lastly, the effect of the ratio of selected clients $p$ is discussed. The variable $p$ is a hyper parameter which controls the proportion of local clients selected to participate in a round of global training. Intuitively, the higher the proportion of clients selected for each round of global training, the better the recommendation accuracy will be. We evaluate the impact of different ratios of selected clients $p \in \{0.2, 0.4, 0.6, 0.8, 1\}$, and we can derive some insightful observations from such results in Fig.~\ref{fig:selected_ratio_lightfedrec}. As the ratio of selected clients $p$ increases from 0.2 to 1, there is generally an upward trend in HR@10 and NDCG@10 of our LightFR, but the improvement tends to stop when $p$ is larger than 0.8 on Movielens-1M, Douban-Movie and Ciao datasets, 0.6 on Filmtrust dataset, respectively. We take the  Movielens-1M dataset as an example, since our method has a similar tendency to the performance impact of $p$ on the above four datasets. As the ratio of selected clients is increased from 0.2 to 0.8, the HR@10 and NDCG@10 have experienced a noticeable rise. This is mainly because, with more aggregation from different clients about the computed gradients or model parameters, the global model can obtain richer information to capture the preferences of users and the attributes of items more accurately. However, as the ratio of selected clients $p$ continues to rise, the quality of recommendation list for users becomes slightly worse since aggregating the gradient information from more different clients means that more noise and unnecessary bias may be introduced. Furthermore, it will take a longer time to train the model since the server needs to wait for more clients to perform local training and aggregate their corresponding training results. Based on these findings, it is crucial for choosing the suitable parameter of $p$ to achieve a good balance between the recommendation quality and the computing efficiency.

\section{Conclusion}

In this work, we propose a lightweight and privacy-preserving federated matrix factorization framework, \textit{LightFR}, which enjoys both fast online inference and economic memory and communication consumption in federated settings. It decentralizes data storage compared with existing hashing based recommender systems. We alleviate the four challenges in designing this framework with learning to hash technique, \textit{i.e.},  the huge memory occupation, the large communication bandwidth and the heavy calculation overheads on the local resource-constrained clients, and privacy protection for parameters transmitting between the server and clients. Besides, we design a federated discrete optimization algorithm between the central server and distributed clients, which can employ collaborative discrete optimization in federated scenarios to produce superior binary user representation on the local client and suitable binary item representations on the server side. Furthermore, we comprehensively discuss the superiority of our model on storage/communication efficiency, inference efficiency, and privacy enhancement from theoretical perspectives. We further conduct extensive experiments and the overall comparing experiments demonstrate that our framework significantly outperforms state-of-the-art FRS methods in terms of recommendation accuracy, resource savings and data privacy. Lastly, detailed sensitivity analysis regarding the hyper parameters further justifies the efficacy of our proposed model integrating learning to hash technique into canonical MF backbone in federated settings.

Despite the effectiveness and efficiency of our LightFR, there are still a few future directions to explore. Firstly, we essentially make a preliminary attempt to introduce the fundamental learning to hash technique into FL framework in recommendation scenario. Therefore, the binary user and item representations could be substantially enhanced by integrating extra side information to obtain more accurate and efficient discrete representations in federated settings~\cite{hash_motivation_2020,semi_dmf_2020}. Secondly, LightFR exclusively designs discrete representation learning on top of vanilla MF model in its current version. In future, with the high flexibility of our proposed framework, we may explore the federated discrete representation learning mechanism for more advanced user modeling algorithms, such as factorization machines~\cite{fm_2010} and graph neural networks~\cite{mf_gnn_2018}, so as to learn more compact and informative binary representations of users and items.

\begin{acks}

The authors would like to thank the anonymous reviewers for their helpful comments and suggestions.

\end{acks}

\bibliographystyle{ACM-Reference-Format}
\bibliography{lightfedrec}

\end{document}